\documentclass[pra,twocolumn,showpacs,amsmath,amssymb]{revtex4-1}
\usepackage{bm}
\usepackage[pdftex]{graphicx}
\usepackage{color}

\begin{document}

\title{Dynamics of two quantized vortices belonging to different components of binary Bose-Einstein condensates in a circular box potential}
\author{Junsik Han$^1$}
\author{Kenichi Kasamatsu$^2$}
\author{Makoto Tsubota$^{1,3,4}$}
\affiliation{$^1$Department of Physics, Osaka City University, 3-3-138 Sugimoto, Sumiyoshi-ku, Osaka 558-8585, Japan}
\affiliation{$^2$Department of Physics, Kindai University, Higashi-Osaka, Osaka 577-8502, Japan}
\affiliation{$^3$The Advanced Research Institute for Natural Science and Technology, Osaka City University, 3-3-138 Sugimoto, Sumiyoshi-ku, Osaka 558-8585, Japan}
\affiliation{$^4$Nambu Yoichiro Institute of Theoretical and Experimental Physics (NITEP), Osaka City University, 3-3-138 Sugimoto, Sumiyoshi-ku, Osaka 558-8585, Japan}
\date{\today}

%%%%%%%%%%%%%%%%%%%%%%%%%%%%%%%%%%%%%%%%%%%%%%%%%%%%%%%%%%%%%%%%%%%%%%%%%%%%%%%%%%%%

\begin{abstract}

This study aims to research on dynamics of two quantized vortices in miscible two-component Bose--Einstein condensates, trapped by a circular box potential by using the Gross--Pitaevskii equation.
We consider a situation in which two vortices belong to different components and they are initially close. 
Their dynamics are significantly different from those of a vortex pair in a single-component condensate.
When two vortices are initially co-located, they are split by dynamical instability or remain co-located depending on their intercomponent interaction.
If the two split vortices have the same sign of circulation, they rotate around each other.
Then, the angular velocity can be given as a function of the distance between two vortices, which is understood through the equations of motion and their interaction.
If the circulations of two close vortices have different signs, however, they move in the same direction initially, then overlap during the dynamics.
This subsequent overlap can be categorized into two types based on the initial distance between them.
The mechanism of this overlapping can be understood from the interaction between the vortex and its image vortex, which makes the two vortices close to each other even if the intrinsic interaction between them is repulsive.

%%% Keywords are not needed any longer. %%%
%%%\kword{keyword1, keyword2, keyword3, \ldots}
%%%

\end{abstract}

\maketitle

%%%%%%%%%%%%%%%%%%%%%%%%%%%%%%%%%%%%%%%%%%%%%%%%%%%%%%%%%%%%%%%%%%%%%%%%%%%%%%%%%%%%

\section{Introduction}\label{sec:Introduction}

The study of the dynamics of quantized vortices is crucial for quantum turbulence (QT) understanding. 
QT is a turbulence in quantum fluids and consists of many tangled quantized vortices. 
A quantized vortex is a topological defect that has a quantized circulation and a core size characterized by the healing length $\xi$.

In dilute atomic Bose--Einstein condensates (BECs), which are typical quantum fluids, the dynamics of QT and vortices have been widely studied \cite{Tsubota13, Barenghi14, Madeira20}.
The BEC system has significant advantages to study the dynamics of quantized vortices.
The first is to realize a quasi-two-dimensional (2D) system by using a tightly confining trap potential along one axis \cite{Kwon14}.
The dynamics of vortices become simple, because a vortex is a singular point in a 2D system \cite{Torres11}.
Second, there are several methods which create vortices in the system.
For example, vortices can be created by the phase imprinting \cite{Matthews99, Leanhardt02}, rotation of the trap potential \cite{Madison01, Tsubota02}, the moving obstacle potential \cite{Sasaki10, Fujimoto11, White12, Neely13, Kwon14, Seo17}, or the flow around the obstacle potential \cite{Frisch92, Reeves13, Reeves15}.
Third, each vortex with the core of the healing length of $\mu {\rm m}$ is visualized using the optical technique.
The sign of the circulation of each vortex can also be confirmed by using the Bragg scattering \cite{Seo17}.

Moreover, quantized vortices in multicomponent BESs have been studied in cold atom systems \cite{Kasamatsu05}.
In two-component BECs, there are two types of interactions between atoms; one is the intracomponent interaction between atoms of the same component, and the other is the intercomponent interaction between atoms of different components.
These two types of interactions between atoms yield the two types of interaction between vortices: intracomponent and intercomponent one.
The asymptotic form of the interaction between well-separated vortices in two-component BECs have been studied \cite{Eto11}, which makes the dynamics of vortices more different from that in one-component BECs.
For example, the formation of a vortex lattice \cite{Kasamatsu03, Aftalion12} and Onsager vortices, which are clusters of same sign vortices \cite{Han18, Han19}.
Recently, the short-range interaction and the associated dynamics of two quantized vortices have also been studied \cite{Kasamatsu16}.
The two vortices approach or leave each other based on their signs of circulation and intercomponent interactions.
Additionally, the dynamics of co-lotated vortices in a harmonic potential are reported \cite{Mithun21}; the co-located vortices are emitted by the moving obstacle potential, and remain co-located when there is no asymmetry of the parameters in the two-component system. 
Here, the co-located vortices mean that the vortices belonging to different components take the same position.
The instability of co-located vortices in a harmonic potential has also been reported \cite{Kuopanportti19}.
We also confirm that co-located vortices are emitted from an  obstacle potential and remain co-located for a long time, as described in the appendix.
However, the following problems are not clear in these previous studies.
First, the condition of splitting of two vortices, which depends on the intercomponent interaction and trapping potential.
Second, the dynamics during the splitting process.
Third, the dynamics after the splitting.
Thus, the understanding of these problems is indispensable for studying multicomponent QT.

In this research, we study the dynamics of vortices in miscible two-component BECs trapped by a circular box potential.
We assume that there are two vortices, where each component has one vortex, and study the dependence of the dynamics on the intercomponent interaction and the signs of the two vortices.
Here, we consider two cases for the initial configuration of two vortices.
First, two vortices are co-located or close to each other at the center of the trapping potential, and the other is that they are separated with an initial distance of the order of $\xi$.
The former was studied by Kuopanportti {\it et al.} \cite{Kuopanportti19} in the harmonic potential, where, the density of condensates was not uniform and the dynamics of vortices were affected by the trapping potential.
The latter was also studied by Kasamatsu {\it et al.}. in a uniform system \cite{Kasamatsu16}, and the results are different from ours because of the box potential.
To understand binary QT, we study the dynamics of two vortices trapped by a circular box potential, where the density is uniform and the effect of the boundary is included.
The circular box potential is introduced by supposing reproduction in an experiment.

We find three characteristic results on the vortex dynamics.
The first is the splitting of the initially co-located vortices at the center of the box potential, which originates from the dynamical instability of the initial state.
Although this dynamical instability of the initially co-located state is studied in the harmonic potential \cite{Kuopanportti19}, we examine whether the dynamical instability also causes the splitting of vortices in the circular box potential, where the density is almost uniform.
Whether the co-located vortices split or remain, depends on the intercomponent interactions between atoms.
The second one is the dynamics of vortices with circulations of same signs.
The two vortices make a circular motion around the center of the potential, and the angular velocity can be calculated using the motion equations for the point-vortex model.
The third one is the dynamics of vortices with circulations of unlike signs.
Then, we focus on the two types of overlapping for a moment.
One occurs when the two vortices are co-located or close to each other, while the other occurs when the two vortices are separated initially, where the interaction between the vortex and its image vortex causes this overlap in both types.
In the former case, the overlapping occurs closer to the center of the potential since the vortices sufficiently overlap even in the initial state and the subtle interaction with the image vortex is enough to cause this overlapping.
In the latter case, two vortices move to near the boundary of the potential, and there is sufficient interaction with the image vortex.
This sufficient interaction with the image vortex is required to cause overlapping in the latter case, and we estimate the region where interaction becomes significant.
These dynamics are significantly different from those of a vortex pair in a single-component BEC.

The rest of the study is organized as follows.
In Sec. \ref{sec:Model}, our numerical model is introduced.
In Sec. \ref{sec:Results}, we show the results of our calculation.
We present the splitting of the vortices in Sec. \ref{sec:Splitting}.
After the splitting, the dynamics depend on the signs of the circulations of the two vortices, and the dynamics of the two vortices with same signs are shown in Sec. \ref{sec:Same_sign}.
In Sec. \ref{sec:Different_sign}, we also describe the dynamics of the two vortices with unlike signs.
Finally, we conclude in Sec. \ref{sec:Conclusions}.

%%%%%%%%%%%%%%%%%%%%%%%%%%%%%%%%%%%%%%%%%%%%%%%%%%%%%%%%%%%%%%%%%%%%%%%%%%%%%%%%%%%%

\section{Model} \label{sec:Model}

Here, we calculate the dynamics of vortices in 2D two-component BECs by using the Gross--Pitaevskii(GP) equations
%%%%%%%%%%
\begin{equation}
	\imath\hbar\frac{\partial}{\partial t}\psi_{j} = \left\{-\frac{\hbar^{2}}{2m_{j}}\nabla^{2} + V + \displaystyle\sum_{j'=1,2} g_{jj'}|\psi_{j'}|^{2} - \mu_{j} \right\} \psi_{j} , \label{eq:GPEs} 
\end{equation}
%%%%%%%%%%
which can quantitatively describe the behavior of dilute atomic BECs at zero temperature \cite{PethickSmith08, Tsatsos16, Tsubota17}.
In these equations, $\psi_{j} = \sqrt{n_{j}({\boldsymbol r},t)}\mathrm{e}^{\imath \theta_{j}({\boldsymbol r},t)}$ is the macroscopic wave function of the $j$-th component ($j = 1, 2$), where $n_{j}({\boldsymbol r},t)$ is the density of the condensate, and $\theta_{j}({\boldsymbol r},t)$ is its phase.
The total particle number of the $j$-th component can be calculated as $N_{j} = \int n_{j}({\boldsymbol r},t) d{\bf r}$.
Here, $m_{j}$ is the mass of the atom, $g_{jj'}$ is the intensity of the atom--atom interaction, and $\mu_{j}$ is the chemical potential.
For simplicity, we choose $N_{1} = N_{2}$, $m_{1} = m_{2} = m$, $g_{11} = g_{22} = g$, and the trap potential is
%%%%%%%%%%
\begin{equation}
	V({\boldsymbol r}) = \left \{
	\begin{array}{l}
	V_{0} \hspace{1pc}(r > R_{0}) \\
	0 \hspace{1.5pc}(r < R_{0})
	\end{array} 
	\right..\label{eq:box_potential}
\end{equation}
%%%%%%%%%%
Here, $r = |{\boldsymbol r}|$, the height of the potential $V_{0} = 1000gn$, and the radius of the potential $R_{0} = 56\xi$, where $n$ is the density of the uniform condensate and the healing length $\xi = \hbar / \sqrt{mgn}$.

In the initial state of the simulations, we imprinted vortices by multiplying the wave function in the $j$-th component by a phase factor $\displaystyle \exp(\imath \theta^{\rm v}_{j})$, where $\theta^{\rm v}_{j}(x,y) = s_{j} \arctan[(y-y_{j})/(x-x_{j})]$.
Here, the coordinates $(x_{j},y_{j})$ and $s_{j}$ refer to the position and sign of the circulation of the vortex of the $j$-th component, respectively.
After imprinting, we solved the wave function using Eq. (\ref{eq:GPEs}) through imaginary time to form the vortex cores \cite{Kasamatsu16}, and then introduced the small random noise at $t=0$ of the real-time simulations.

In the calculation, Eq. (\ref{eq:GPEs}) was made dimensionless by the length, time, and energy scales $\xi$, $\tau = \hbar / gn$, and $\epsilon = gn$.
The dimensionless wave function and intercomponent interaction are defined as $\tilde{\psi}_{j}({\boldsymbol r},t) = \displaystyle \frac{1}{\sqrt{n}} \psi_{j}({\boldsymbol r},t)$ and $\tilde{g}_{12} = g_{12} / g$.
We performed these numerical calculations by using the Fourier spectrum method and the fourth-order Runge-Kutta method.
The space and time were divided by $\Delta x = 0.25\xi$ and $\Delta t = 1.0 \times 10^{-3} \tau$.
In the following part, we used the dimensionless quantities without tilde.

Using the form $\psi_{j} = \sqrt{n_{j}({\boldsymbol r},t)}\mathrm{e}^{\imath \theta_{j}({\boldsymbol r},t)}$, the kinetic energy $E^{\rm kin}_{j}$, the intracomponent interaction energy $E^{\rm int}_{j}$, and the intercomponent interaction energy $E^{\rm int}_{12}$ can be calculated as
%%%%%%%%%%
\begin{equation}
	E^{\rm kin}_{j}(t) = \frac{1}{2}\int \left | \nabla \left\{ \sqrt{n_{j}({\boldsymbol r},t)} e^{\imath \theta_{j}({\boldsymbol r},t)} \right\} \right |^{2} d{\boldsymbol r}, \label{eq:Ekin}
\end{equation} 
%%%%%%%%%%
%%%%%%%%%%
\begin{equation}
	E^{\rm int}_{j}(t) = \frac{g}{2}\int n_{j}({\boldsymbol r},t)^{2} d{\boldsymbol r}, \label{eq:Eint}
\end{equation} 
%%%%%%%%%% 
and
%%%%%%%%%%
\begin{equation} 
	E^{\rm int}_{12}(t) = g_{12}\int n_{1}({\boldsymbol r},t)n_{2}({\boldsymbol r},t) d{\boldsymbol r}, \label{eq:Eint12}
\end{equation} 
%%%%%%%%%%
respectively.
We defined the total kinetic and interaction energies as 
%%%%%%%%%%
\begin{equation}
	E^{\rm kin}(t) = E^{\rm kin}_{1}(t) + E^{\rm kin}_{2}(t), \label{eq:Ekin_tot}
\end{equation} 
%%%%%%%%%% 
and 
%%%%%%%%%%
\begin{equation}
	E^{\rm int}(t) = E^{\rm int}_{1}(t) + E^{\rm int}_{2}(t) + E^{\rm int}_{12}(t). \label{eq:Eint_tot}
\end{equation} 
%%%%%%%%%% 
Using these equations, the total energy $E^{\rm tot}$ is defined as
%%%%%%%%%%
\begin{equation}
	E^{\rm tot}(t) = E^{\rm kin}(t) + E^{\rm int}(t) + E^{\rm pot}(t) = \int \epsilon^{\rm tot}({\boldsymbol r}, t) d{\bf r}, \label{eq:Etot}
\end{equation} 
%%%%%%%%%%
where
%%%%%%%%%%
\begin{equation}
	E^{\rm pot}(t) = E^{\rm pot}_{1}(t) + E^{\rm pot}_{2}(t), \label{eq:Epot_tot}
\end{equation} 
%%%%%%%%%% 
%%%%%%%%%%
\begin{equation}
	E^{\rm pot}_{j}(t) = \int V({\boldsymbol r}) n_{j}({\boldsymbol r},t) d{\boldsymbol r}, \label{eq:Epot}
\end{equation} 
%%%%%%%%%% 
and $\epsilon^{\rm tot}({\boldsymbol r}, t)$ is the density of the total energy.

%%%%%%%%%%%%%%%%%%%%%%%%%%%%%%%%%%%%%%%%%%%%%%%%%%%%%%%%%%%%%%%%%%%%%%%%%%%%%%%%%%%%

\section{Results} \label{sec:Results}

We therefore write the coordinates of the vortex, the distance of the vortex of the $j$-th component from the center of the potential, and the distance between the vortices in two components as $(x_{j}(t), y_{j}(t))$, $r_{j}(t)$, and $d(t)$, respectively.

%%%%%%%%%%%%%%%%%%%%%%%%%%%%%%%%%%%%%%%%

\subsection{Splitting of vortices} \label{sec:Splitting}

In this section, we show the splitting of two co-located vortices, one in component 1 and the other in component 2. 
These two vortices were initially imprinted at the center of the potential, and thus, $(x_{j}(0), y_{j}(0)) = (0, 0)$.

To study the dynamical instability of the initial state, we assume that the wave function can be written using the polar coordinates $(r, \varphi)$ as
%%%%%%%%%%
\begin{equation}
	\psi_{j}(r, \varphi, t) = e^{-\imath \mu_{j} t} e^{\imath s_{j} \varphi} [f_{j}(r) + \chi_{j}(r, \varphi, t)]. \label{eq:polar_wf_chi}
\end{equation} 
%%%%%%%%%% 
Here, $f_{j}$ is the density profile of the stationary solution of Eq. (\ref{eq:GPEs}), and $\chi_{j}$ is a small deviation from $f_{j}$.
By substituting Eq. (\ref{eq:polar_wf_chi}) into Eq. (\ref{eq:GPEs}), ignoring the second- and third-order terms in $\chi$ and assuming that the form of the solution is 
%%%%%%%%%%
\begin{equation}
	\chi_{j}(r, \varphi, t) = \sum_{q} \sum_{l} [ u^{(l)}_{q, j}(r) e^{\imath l \varphi - \imath \omega^{(l)}_{q} t} + v^{(l) *}_{q, j}(r) e^{-\imath l \varphi + \imath \omega^{(l) *}_{q} t} ], \label{eq:chi}
\end{equation} 
%%%%%%%%%% 
we obtained the Bogoliubov-de Gennes (BdG) equations
%%%%%%%%%%
\begin{equation}
	\mathcal{B}^{(l)} {\boldsymbol w}^{(l)}_{q}(r) = \omega^{(l)}_{q} {\boldsymbol w}^{(l)}_{q}(r), \label{eq:Bogoliubov}
\end{equation} 
%%%%%%%%%% 
where 
%%%%%%%%%%
\begin{equation}
	{\boldsymbol w}^{(l)}_{q}(r) = \left(u^{(l)}_{q, 1}(r), u^{(l)}_{q, 2}(r), v^{(l)}_{q, 1}(r), v^{(l)}_{q, 2}(r) \right)^{\rm T}, \label{eq:Bogoliubov_w}
\end{equation} 
%%%%%%%%%% 
and
%%%%%%%%%%
\begin{equation}
	\mathcal{B}^{(l)} = 
	\begin{pmatrix}
		\mathcal{D}^{(l + s_{1})}_{1} & g_{12} f_{1} f^{*}_{2} & g f^{2}_{1} & g_{12} f_{1} f_{2} \\
		g_{12} f^{*}_{1} f_{2} & \mathcal{D}^{(l + s_{2})}_{2} & g_{12} f_{1} f_{2} & g f^{2}_{2} \\
		-g(f^{*}_{1})^{2} & -g_{12} f^{*}_{1} f^{*}_{2} & -\mathcal{D}^{(l - s_{1})}_{1} & -g_{12} f^{*}_{1} f_{2} \\
		-g_{12} f^{*}_{1} f^{*}_{2} & -g(f^{*}_{2})^{2} & -g_{12} f_{1} f^{*}_{2} & -\mathcal{D}^{(l - s_{2})}_{2} \\
	\end{pmatrix}
	. \label{eq:Bogoliubov_B}
\end{equation} 
%%%%%%%%%% 
The integer $l$ is the angular momentum of the excitation with respect to the condensate, $q$ is an index for the different eigensolutions with the same $l$, and the operators $\mathcal{D}^{(l \pm s_{j})}_{j}$ are written as
%%%%%%%%%%
\begin{eqnarray}
	\mathcal{D}^{(l \pm s_{j})}_{j} = &-&\frac{1}{2} \left \{ \frac{1}{r} \frac{d}{dr} \left( r\frac{d}{dr} - \frac{(l \pm s_{j})^{2}}{r^{2}} \right) \right \} \nonumber \\ 
	&+& V -\mu_{j} + 2g|f_{j}|^{2} + g_{12}|f_{j'}|^{2} \ (j' \neq j). \ \label{eq:Bogoliubov_D}
\end{eqnarray}
%%%%%%%%%% 

\begin{center}
\begin{figure}[h]
	\includegraphics[width=18pc, height=12pc]{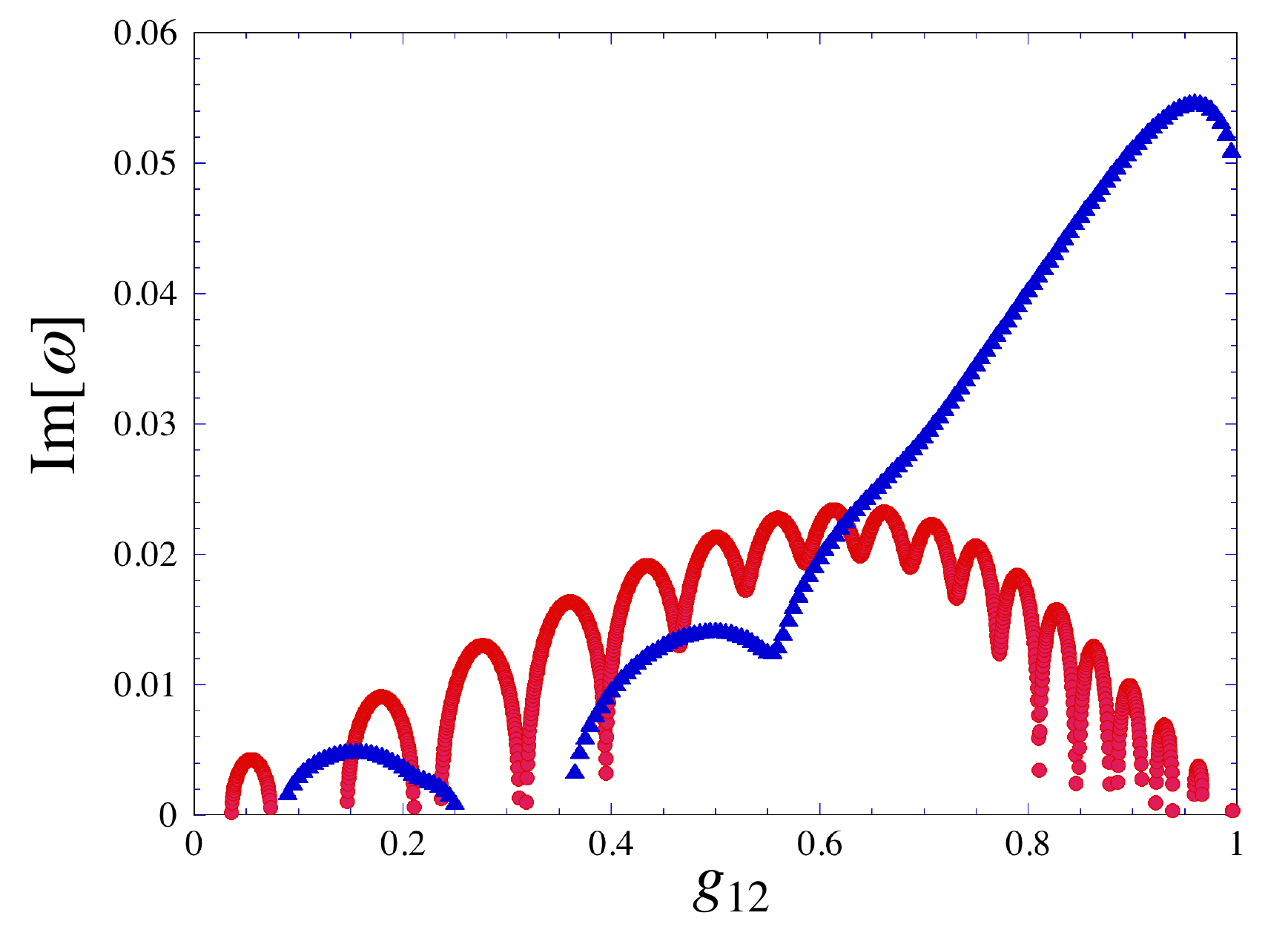}
\caption{\label{f1} 
The imaginary part of the frequency $\omega$ of the eigenmode for the angular momentum quantum number $l = 1$ is presented as a function of the intercomponent interaction.
Here, the stationary state takes the vortices at $(x_{j}(0), y_{j}(0)) = (0,0)$, the red circles represent $s_{1} = s_{2} = +1$, and the blue triangles represent $s_{1} = +1$ and $s_{2} = -1$.
}
\end{figure}
\end{center}

From the BdG equation (Eqs. (\ref{eq:Bogoliubov})), the dynamical instability condition can be obtained.
If there is an integer $l$ for which $\omega^{(l)}_{q}$ includes at least one eigenfrequency that has a positive imaginary part ${\rm Im}[\omega^{(l)}_{q}] > 0$, the amplitude of the excitation mode grows exponentially with time, and thus, the state is dynamically unstable.
Figure \ref{f1} is the stability diagram of the state where the signs of the vortex circulations are $s_{1} = s_{2} = +1$ (red circle) and $s_{1} = +1$, $s_{2} = -1$ (blue triangle) for their initial position $(x_{j}(0), y_{j}(0)) = (0, 0)$.
We determine that ${\rm Im} [\omega_q^{(l)}]$ appears for the eigenmodes with $l=\pm 1$ and a certain value of $q$.

%%%%%%%%%%%%%%%%%%%%
%%%%%%%%%%%%%%%%%%%%

\begin{center}
\begin{figure}[h]
	\includegraphics[width=20pc, height=13.158pc]{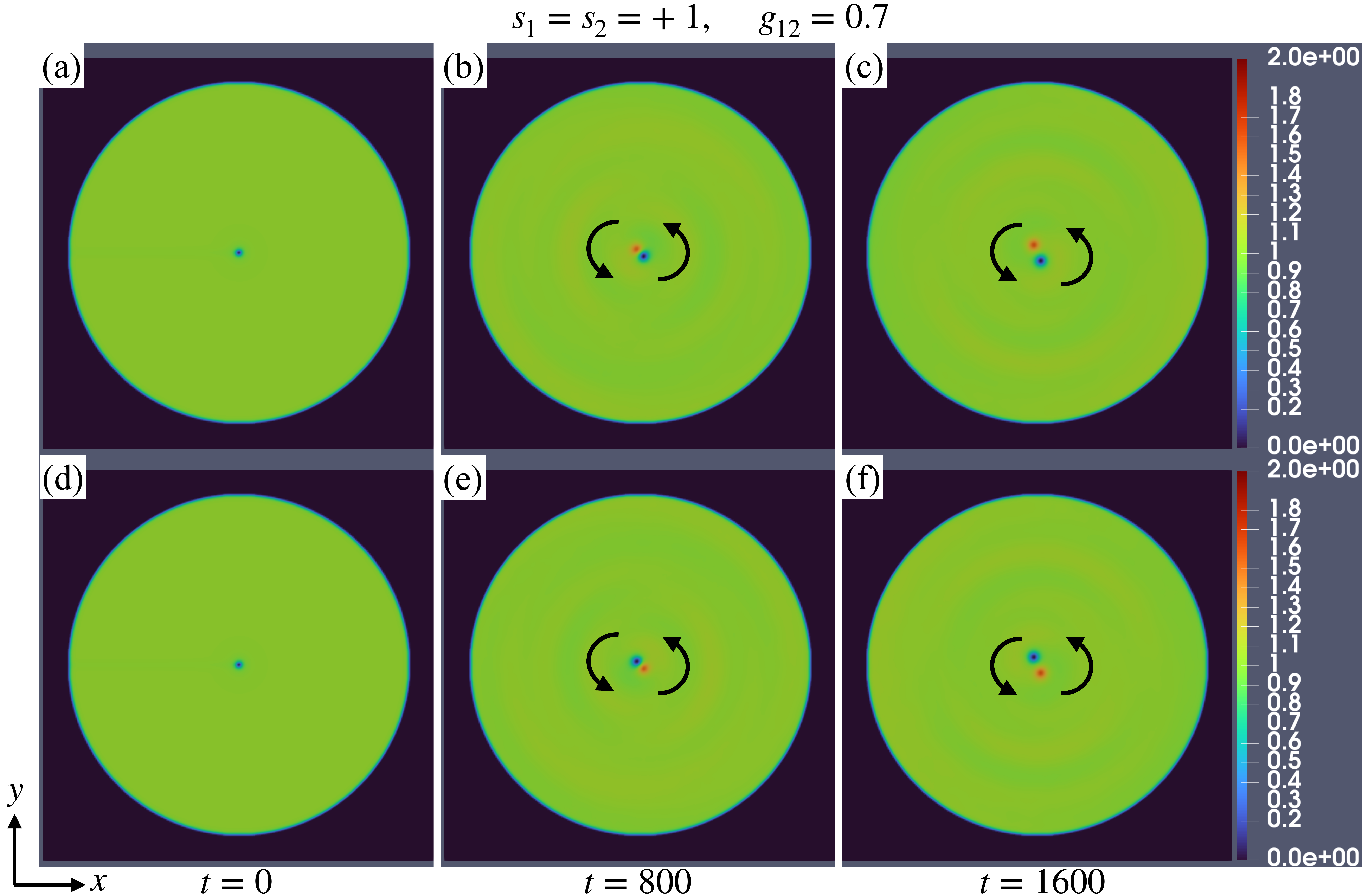}
\caption{\label{f2} 
The densities of (a) - (c) component 1, and (d) - (f) component 2, for $s_{1} = s_{2} = +1$, and $g_{12}=0.7$.
The vortex of each component was initially imprinted at the center of the potential $(x_{j}(0), y_{j}(0)) = (0, 0)$.
The black arrows indicate the direction of the moving of vortices.}
\end{figure}
\end{center}

The stable state is expected to appear in the narrow region around the $g_{12} = 0$ and $g_{12} = 0.1$ for $s_{1} = s_{2} = +1$.
When $g_{12} = 0.1$, the two vortices remained co-located, although we followed the dynamics until $t =1600$.
However, as shown in Fig. \ref{f2}, the vortices split when $g_{12}=0.7$.
Here, we define the split of vortices as the condition $d > 2$ (in units of $\xi$) because the size of the vortex core was characterized by $\xi$.
The densities at $t = 0$, $t=800$, and $t=1600$ are shown in Fig. \ref{f2}, where, the blue points indicate the low-density vortex cores and the red points indicate the high density hump, which comes from the filling of a vortex core in the other component because of the repulsive interaction \cite{Eto11}.
These results are consistent with the stability diagram (Fig. \ref{f1}).

%%%%%%%%%%%%%%%%%%%%
%%%%%%%%%%%%%%%%%%%%

\begin{center}
\begin{figure}[h]
	\includegraphics[width=20pc, height=13.684pc]{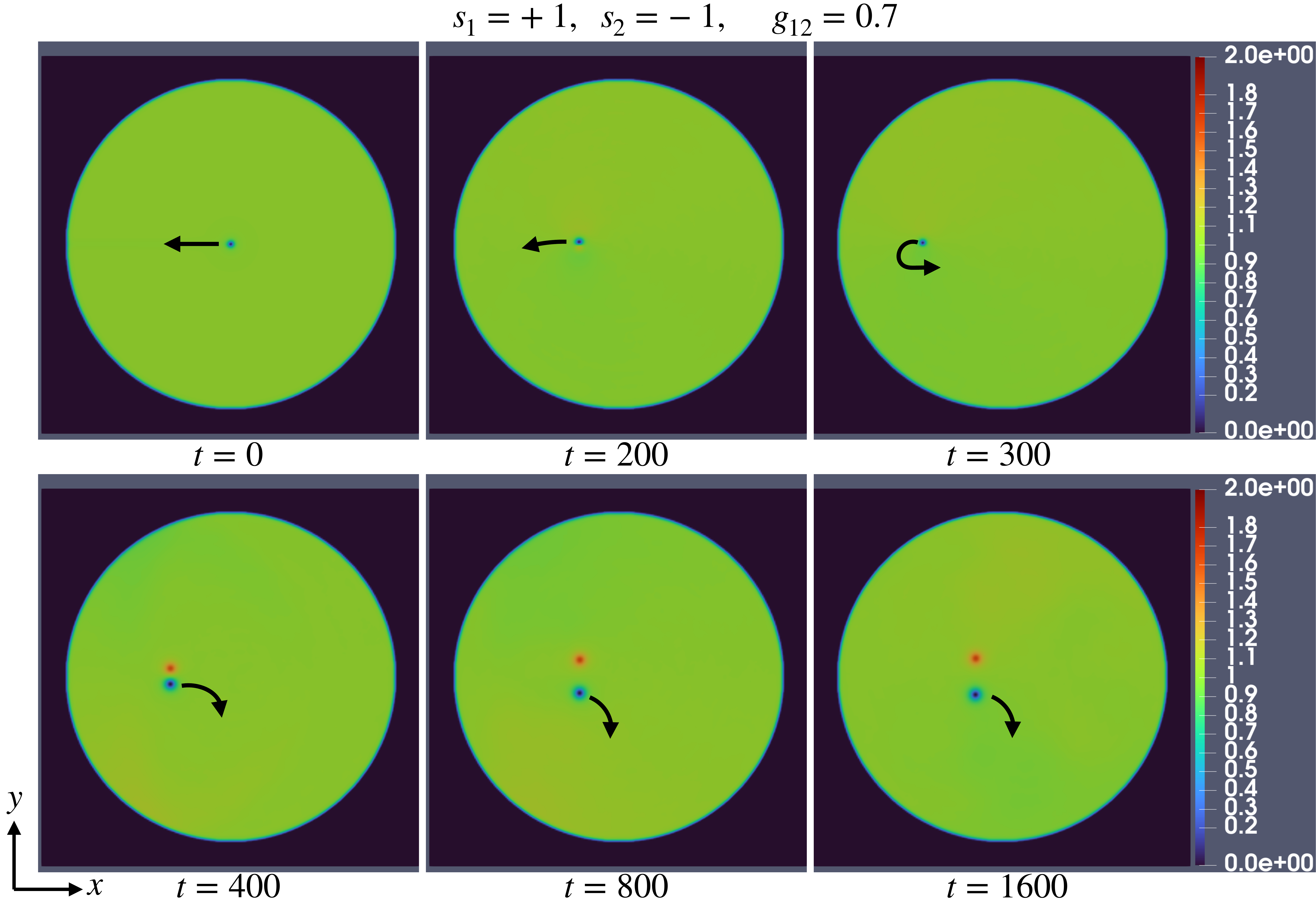}
\caption{\label{f3} 
The density of component 1 for $s_{1} = + 1$, $s_{2} = -1$, and $g_{12}=0.7$.
The vortex of each component was initially imprinted at the center of the potential $(x_{j}(0), y_{j}(0)) = (0, 0)$.
The black arrows indicate the direction of the moving of vortex of component 1. 
The position of the vortex in component 2 was symmetric to that of the vortex in component 1 with  respect to the $x$ axis.}
\end{figure}
\end{center}

The stability diagram of $s_{1} = + 1$, $s_{2} = -1$ is also shown in Fig. \ref{f1}.
Under these conditions, a stable region also appeared around $g_{12} = 0$ and $g_{12} = 0.3$.
When we calculated for $g_{12} = 0.3$, the vortices remained co-located.
Conversely, the vortices split for $g_{12} = 0.7$, as shown in Fig. \ref{f3}.
The details of the vortex dynamics after splitting are described in Sec. \ref{sec:Different_sign}.

%%%%%%%%%%%%%%%%%%%%
%%%%%%%%%%%%%%%%%%%%

%%%%%%%%%%%%%%%%%%%%%%%%%%%%%%%%%%%%%%%%

\subsection{The dynamics of same sign vortices} \label{sec:Same_sign}

If the circulations of two vortices have the same sign, they make a circular motion around the center of the potential after splitting (Fig. \ref{f2}).
Then, the direction of rotation corresponds to that of their circulation.
These results are consistent with that of previous studies \cite{Kasamatsu16, Kuopanportti19}.

However, after splitting, we determined the dynamics of the vortices beyond the scope of previous studies \cite{Kasamatsu16, Kuopanportti19}.
Kasamatsu {\it et al.} derived the equations of motion of two-point vortices by considering the effect of filling one component into the vortex core of the other component \cite{Kasamatsu16}.
The motion equations can be written as
%%%%%%%%%%
\begin{equation}
\begin{split}
 	2\pi s_{1} \dot{y}_{1} &= -\frac{\partial V_{12}}{\partial x_{1}}, \ \ \ \ -2\pi s_{1} \dot{x}_{1} = -\frac{\partial V_{12}}{\partial y_{1}}, \\
	2\pi s_{2} \dot{y}_{2} &= -\frac{\partial V_{12}}{\partial x_{2}}, \ \ \ \ -2\pi s_{2} \dot{x}_{2} = -\frac{\partial V_{12}}{\partial y_{2}}. 
\label{eq:motion}
\end{split}
\end{equation}
%%%%%%%%%% 
Here, $V_{12}$ is the potential of the intercomponent interaction between vortices, taking the asymptotic form \cite{Eto11}
%%%%%%%%%%
\begin{equation}
	V^{\rm asy}_{12} = \frac{4 \pi g_{12}}{1 - g_{12}} \frac{\ln(d/2 \xi_{ \rm cut})}{d^{2}}, \label{eq:V_12}
\end{equation}
%%%%%%%%%%
where $\xi_{\rm cut}$ is the cutoff length of the order of healing length $\xi$.
This asymptotic form is applicable when $d/2 > \xi_{\rm cut}$.

\begin{center}
\begin{figure}[h]
	\includegraphics[width=15pc, height=10.23pc]{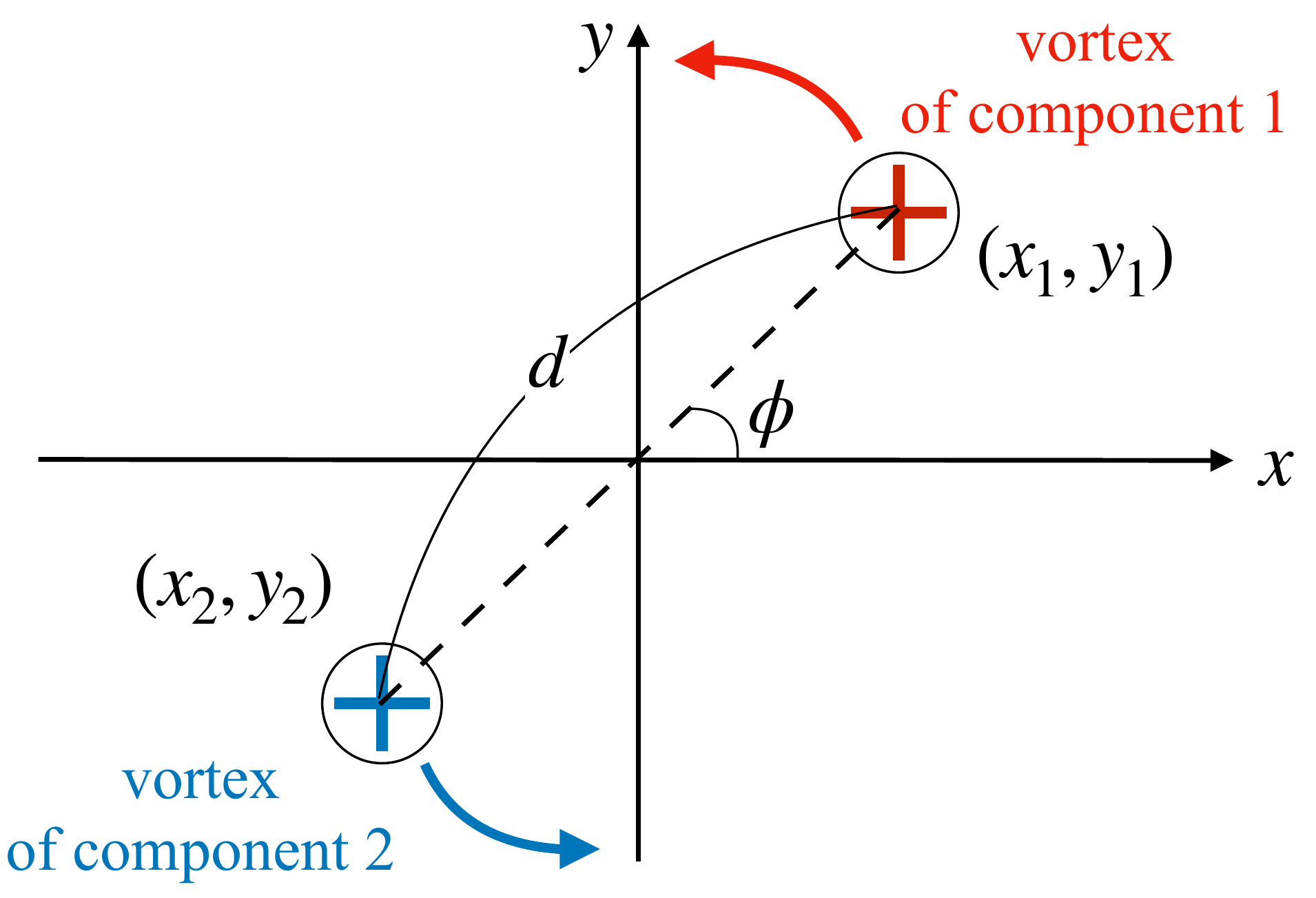}
\caption{\label{f4} 
The schema of rotating vortices.
The red and blue vortices are the vortex of component 1 and 2 respectively.
The distance between them is $d$, and the angle between the position vector of the vortex of component 1 and the $x$-axis is $\phi$.
Here, two vortices rotate around the origin, and thus their positions have the relation $(x_{2}(t), y_{2}(t)) = (-x_{1}(t), -y_{1}(t))$.}
\end{figure}
\end{center}

The distance $d(t)$ between vortices is constant as long as the dynamics follow Eqs.~(\ref{eq:motion}):
The direction of a circular motion of the two vortices can be realized by Eqs.~(\ref{eq:motion}):
Here, we define $\phi(t)$ as the angle between the position vector of the vortex of component 1 and the $x$-axis (Fig. \ref{f4}).
Using Eqs. (\ref{eq:motion}), the angular velocity of the vortex motion $\frac{\rm d}{{\rm d}t}\phi(t)$ is obtained as a function of $d$.
%%%%%%%%%%
\begin{equation}
	\frac{\rm d}{{\rm d} t} \phi = \frac{32g_{12}}{1-g_{12}}\left \{ \ln \left(\frac{d}{2\xi_{\rm cut}} \right) - 1 \right \} \frac{1}{d^{4}}. \label{eq:phi_dif}
\end{equation}
%%%%%%%%%%

%%%%%%%%%%%%%%%%%%%%
%%%%%%%%%%%%%%%%%%%%

\begin{center}
\begin{figure}[h]
	\includegraphics[width=20pc, height=20pc]{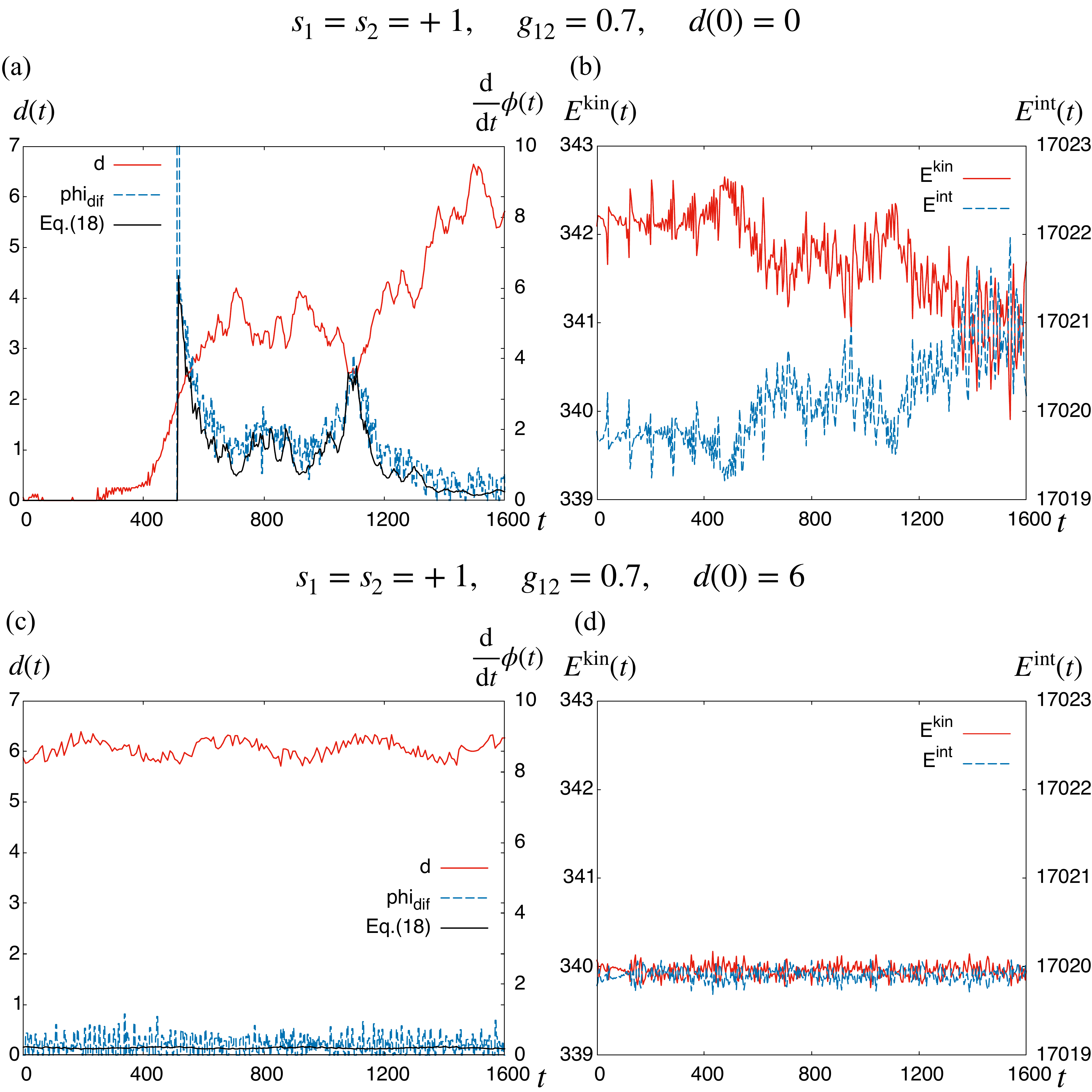}
\caption{\label{f5} 
The time development of (a), (c) $d(t)$, and $\frac{\rm d}{{\rm d}t}\phi(t)$, and (b), (d) $E_{\rm kin}(t)$ and $E_{\rm int}(t)$ for $s_{1} = s_{2} = +1$, $g_{12}=0.7$, and $R_{0} = 56$.
Here, (a) and (b) are the results for $d(0) = 0$, and (c) and (d) are results for $d(0) = 6$.
$d(t)$ is the distance between two vortices, and $\phi(t)$ is the angle between the position vector of the vortex of component 1 and the $x$-axis (Fig. \ref{f4}).
The solid black curves in (a) and (c) show the results of Eq. (\ref{eq:phi_dif}) after substituting $d(t)$ of the numerical simulation (denoted by the red curves).
In (b) and (d), the total kinetic energy $E^{\rm int}(t)$ and interaction energy $E^{\rm int}(t)$ are calculated using Eqs. (\ref{eq:Ekin}) - (\ref{eq:Eint_tot}).
}
\end{figure}
\end{center}

The time developments of $d(t)$ and $\frac{\rm d}{{\rm d}t}\phi(t)$ calculated by the simulation of the GP equations are shown in Fig. \ref{f5} (a).
The distance $d(t)$ rapidly increases when splitting occurs at $t \simeq 400$, and maintains a steady behavior once until $t \simeq 1000$.
After $t \simeq 1000$, $d(t)$ again increases, where the dynamics are affected by the geometry and size of the trap potential.
In these processes, vortices continue to make a circular motion, and the angular velocity also changes (Fig. \ref{f5} (a), blue line).
This angular velocity agrees with Eq. (\ref{eq:phi_dif}), calculated using the obtained $d(t)$ (Fig. \ref{f5} (a), black line). 
Here, we only show that $\frac{\rm d}{{\rm d}t}\phi(t)$ for $d(t) \geq 2$ because the vortices almost overlap for $d(t) < 2$.

%%%%%%%%%%%%%%%%%%%%
%%%%%%%%%%%%%%%%%%%%

\begin{center}
\begin{figure}[h]
	\includegraphics[width=15pc, height=12pc]{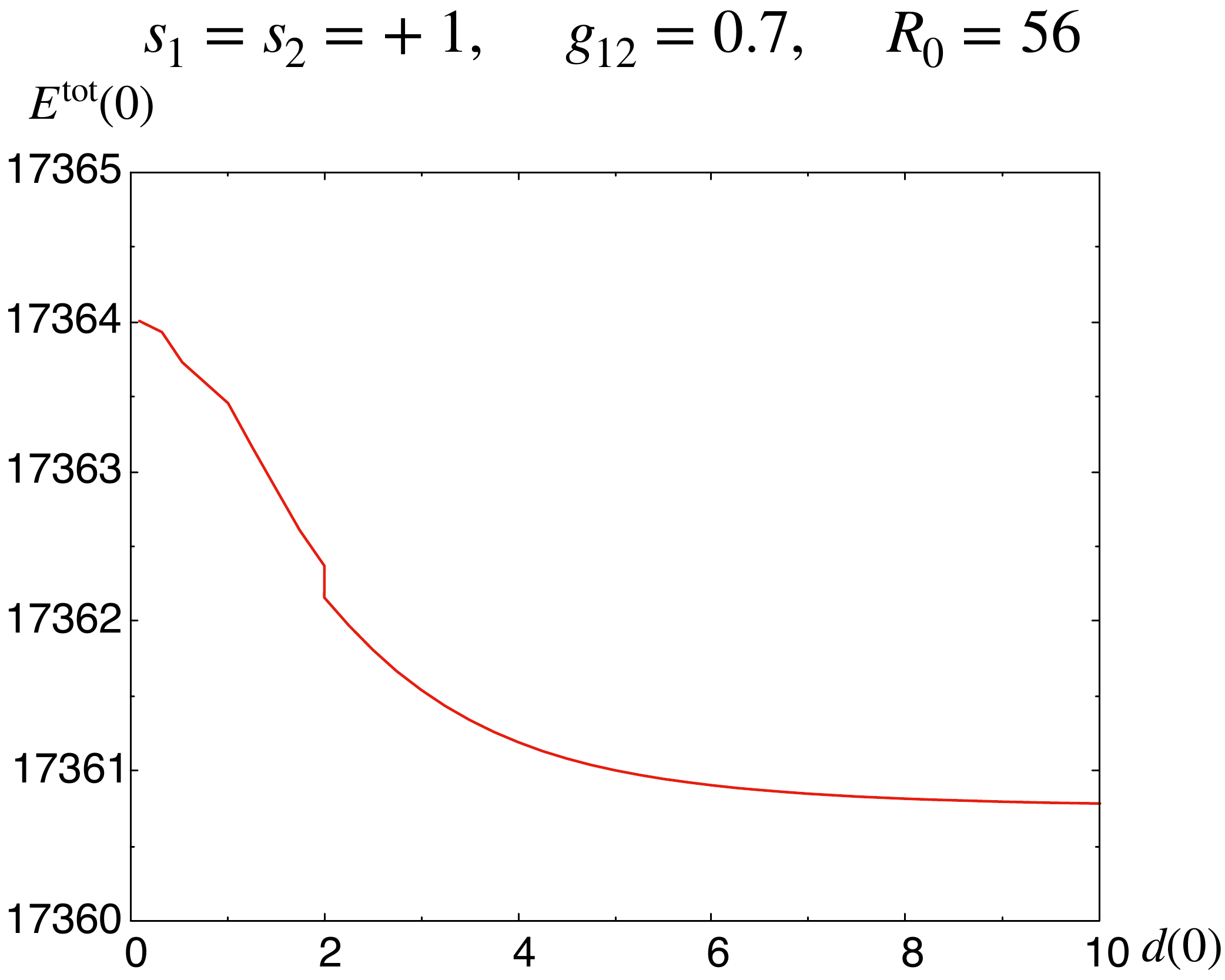}
\caption{\label{f6} 
The initial total energy $E^{\rm tot}(0)$ is shown as a function of the initial distance between the vortices $d(0)$ for $s_{1} = s_{2} = +1$, $g_{12}=0.7$, and $R_{0} = 56$.
The total energy is defined by Eqs. (\ref{eq:Etot}).}
\end{figure}
\end{center}

This nontrivial evolution of the angular velocity can be explained by the temporal change in distance $d(t)$ (Eq. (\ref{eq:phi_dif})) .
Because $d(t)$ should be constant when the motions of the two vortices are governed by Eqs. (\ref{eq:motion}), the change in $d(t)$ cannot be explained only using Eqs.~(\ref{eq:motion}):
To clarify the reason, we also calculated the vortex dynamics with the initial condition $d(0) = 6$.
Then, the changes in $d(t)$ and $\frac{\rm d}{{\rm d}t}\phi(t)$ became very small, as shown in Fig. \ref{f5} (c).
From these results, the change in $d(t)$ can be attributed to the splitting of vortices owing to the dynamical instability of the co-located state and the excess energy that appears when the vortices split.
This excess energy can be estimated from the total energy of the initial state $E^{\rm tot}_{0}$, which is determined by the positions of the two vortices.
Figure \ref{f6} shows the $E^{\rm tot}(0)$ as a function of $d(0)$; $E^{\rm tot}(0)$ is the decreasing function for $d(0)$ because the intercomponent interaction between vortices are repulsive under the condition $g_{12} > 0$.
Here, the decrease in $E^{\rm tot}(0)$ is clear for $d \lesssim 5$; however, $E^{\rm tot}(0)$ is almost constant for $d(0) > 5$.
When the splitting of vortices occurs, the excess energy corresponds to the difference between $E^{\rm tot}(0)$ of $d(0) = 0$, and that of $d(0) > 0$ is emitted as phonon excitation.
However, the excess energy is hardly emitted if the two vortices initially separated sufficiently ($d(0) > 5$).
The change in $d(t)$ after the splitting could be owing to the interactions between the phonons and the vortices.
When $d(t)$ increases and excess energy is emitted, the kinetic energy $E_{\rm kin}(t)$ is transferred into the interaction energy $E_{\rm int}(t)$ (Fig. \ref{f5} (b), $t \simeq 400$).
However, both $E_{\rm kin}(t)$ and $E_{\rm int}(t)$ are almost constant when $d(0) = 6$ (Fig. \ref{f5} (d)).

%%%%%%%%%%%%%%%%%%%%
%%%%%%%%%%%%%%%%%%%%

%%%%%%%%%%%%%%%%%%%%%%%%%%%%%%%%%%%%%%%%

\subsection{The dynamics of different sign vortices} \label{sec:Different_sign}

\begin{center}
\begin{figure}[h]
	\includegraphics[width=18pc, height=15pc]{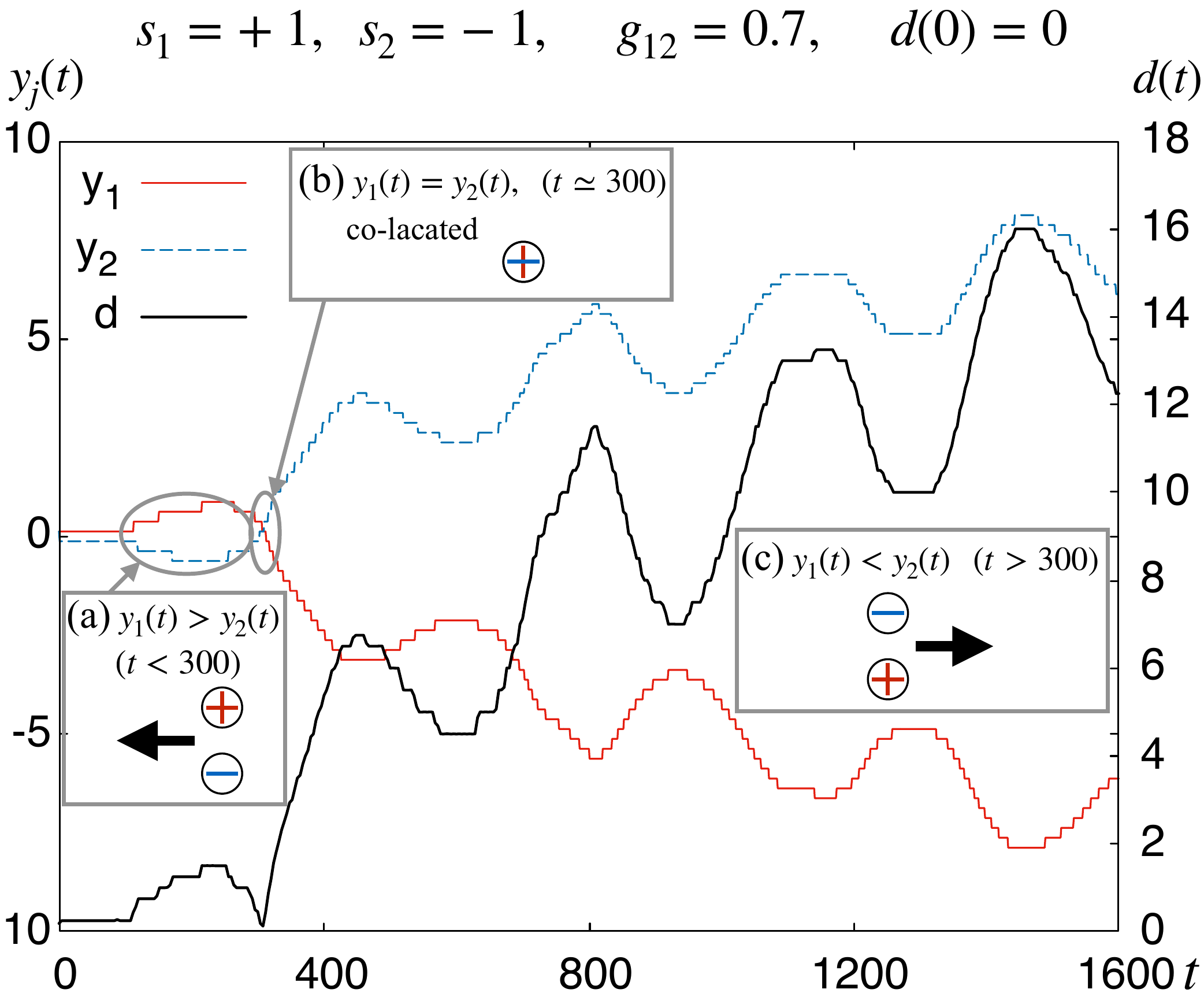}
\caption{\label{f7} 
The time development of $y_{1}(t)$, $y_{2}(t)$, and $d(t)$ for $s_{1} = +1$, $s_{2} = -1$, and $g_{12}=0.7$, started from the co-located vortices at the origin.
The inset shows the schema of (a), in which two slightly shifted vortices move in the same direction for $t < 300$, (b) the instantaneous overlapping of the vortices at $t \simeq 300$, and (c) the positional and motional reverse of the two vortices for $t > 300$.
}
\end{figure}
\end{center}

Figure \ref{f7} shows the time development of $d(t)$ when the two vortices with different signs are co-located initially.
Owing to the dynamic instability of the co-located state (Fig. \ref{f1}), the two vortices slightly shifted each other and moved to the same $-x$ axis direction, perpendicular to the dipole of the two vortices (Fig. \ref{f3}, $t = 0 \sim 200$) with keeping $d(t) < 2$ (Fig. \ref{f7}).
The direction of motion is opposite to that of a vortex pair in a one-component system.
During this process, the positions of the vortices were sustained as $y_{1} > y_{2}$ (Fig. \ref{f7}, inset (a)), and the direction of motion can be realized by Eqs.~(\ref{eq:motion}):
Subsequently, $t \simeq 300$, experienced an instantaneous overlap with $d(t) = 0$ (Fig. \ref{f7}, inset (b)).
Subsequently, the positions of the two vortices were exchanged ($y_{1} < y_{2}$, Fig. \ref{f7}, inset (c)) and evolved as with $d(t) > 2$.
Eventually, they moved in the $+x$-axis direction with increasing the distance $d(t)$.

%%%%%%%%%%%%%%%%%%%%
%%%%%%%%%%%%%%%%%%%%

\begin{center}
\begin{figure}[h]
	\includegraphics[width=20pc, height=16pc]{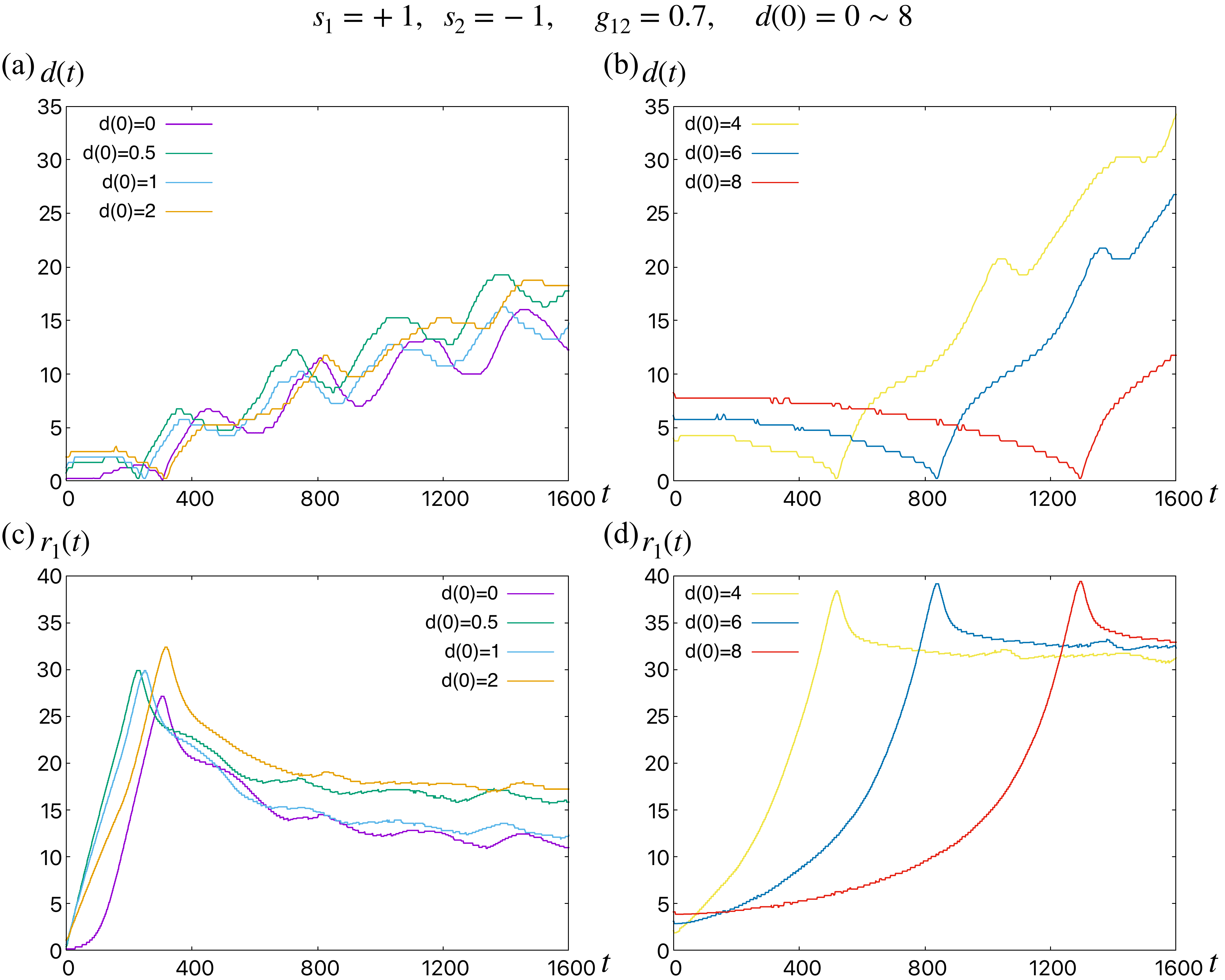}
\caption{\label{f8}
The time development of (a) the distance between vortices $d(t)$ for $d(0) = 0, \ 0.5, \ 1, \ 2$, (b) $d(t)$ for $d(0) = 4, \ 6, \ 8$; (c) the distance of the vortex of component 1 from the center of potential $r_{1}(t)$ for $d(0) = 0, \ 0.5, \ 1, \ 2$, and (d) $r_{1}(t)$ for $d(0) = 4, \ 6, \ 8$, for $s_{1} = + 1$, $s_{2} = -1$, and $g_{12}=0.7$.
The initial positions of the vortices were $(x_{1}(0), y_{1}(0)) = (0, d(0)/2)$ and $(x_{2}(0), y_{2}(0)) = (0, -d(0)/2)$.}
\end{figure}
\end{center}

The intercomponent interaction between the vortices is now repulsive because $g_{12} = 0.7 > 0$.
Then, their overlapping at $t \simeq 300$ is contrary to intuition.
To consider this problem, we calculate the dynamics with the condition $d(0) = 0, \ 0.5, \ 1, \ 2, \ 4, \ 6, \ 8$.
Figure \ref{f8} shows the time development of $d(t)$ and the distance $r_{1}(t)$ between the position of the vortex of component 1 and the center of the potential.
Notably, in all these cases, is that $r_{1}(t)$ initially increases and then decreases after the two vortices overlap and reverse.
From these developments, we can categorize the overlapping into two types, depending on the initial distance $d(0) \leq 2$ or $d(0) > 2$.
For $d(0) \leq 2$, the two vortices are co-located or close to each other.
Then, the time until their overlap, which is defined as $T_{\rm o}$, is almost independent of $d(0)$, and $d(t)$ hardly increases until $t = T_{\rm o}$ (Fig. \ref{f8} (a)).
For $d(0) > 2$, the two vortices are initially sufficiently separated.
In this case, $T_{\rm o}$ increased significantly with $d(0)$ (Fig. \ref{f8} (b)), however the overlapping occured almost at the same distance $r_{1}(T_{\rm o}) \simeq 40$ (Fig. \ref{f8} (d)).
This is because that the initial velocity of the two vortices decreased with $d(0)$, which requires more time for them to reach $r_{1}(T_{\rm o}) \simeq 40 $.
The locations where the overlapping occurred were different as $r_{1}(T_{\rm o}) \simeq 30$ for $d(0) \leq 2$, whereas $r_{1}(T_{\rm o}) \simeq 40$ for $d(0) > 2$.
Because $r_{1}(T_{\rm o}) \simeq 40 $ is close to the radius $R_{0} = 56$ of the potential, the overlapping events occur near its boundary. 

%%%%%%%%%%%%%%%%%%%%
%%%%%%%%%%%%%%%%%%%%

\begin{center}
\begin{figure}[h]
	\includegraphics[width=15pc, height=11.25pc]{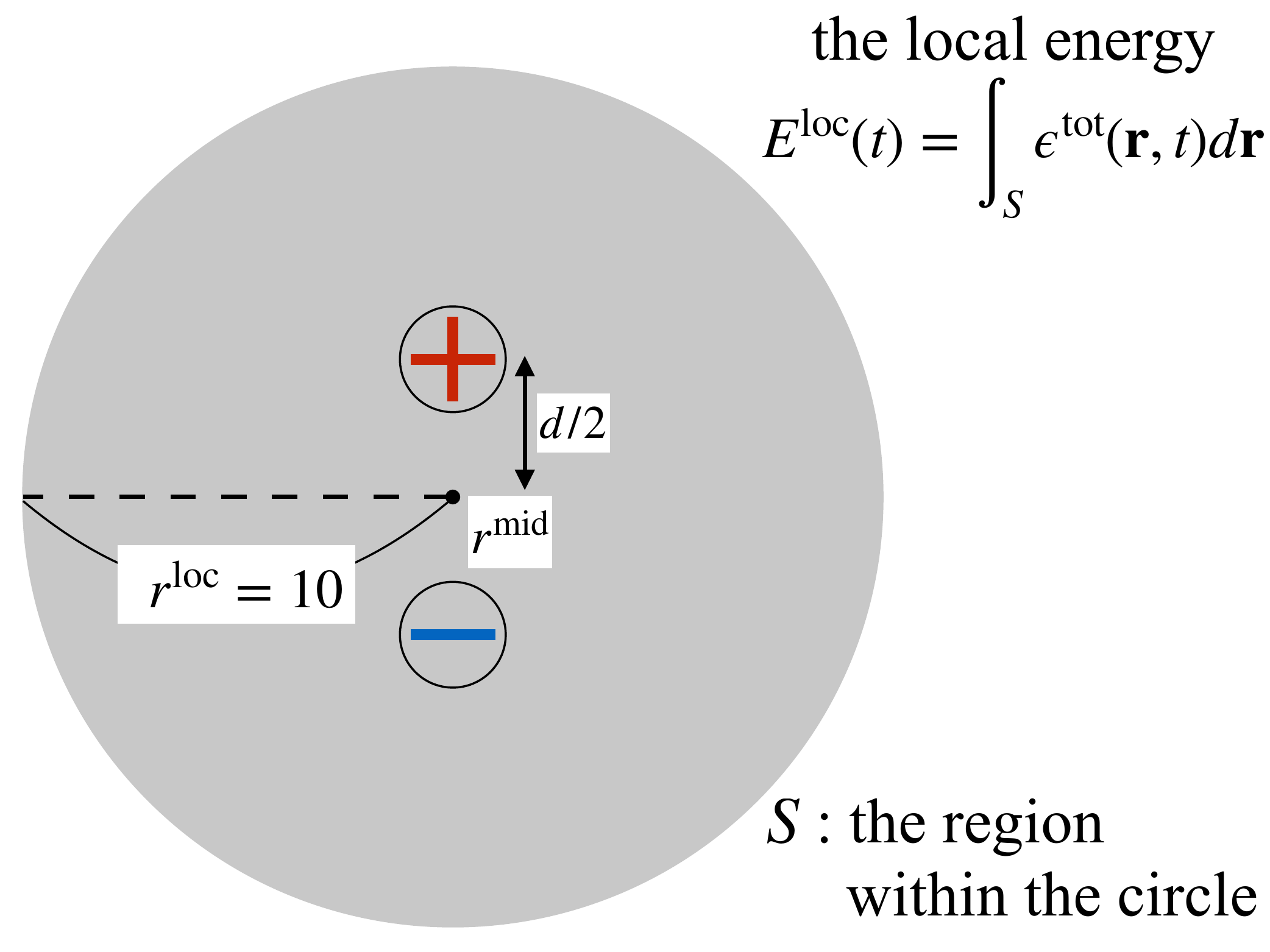}
\caption{\label{f9} 
The schema of the region $S$ within the circle (gray region), whose center and radius are the centers of two vortices $r^{\rm mid}$ and $r^{\rm loc} = 10$.}
\end{figure}
\end{center}

\begin{center}
\begin{figure}[h]
	\includegraphics[width=20pc, height=8.33pc]{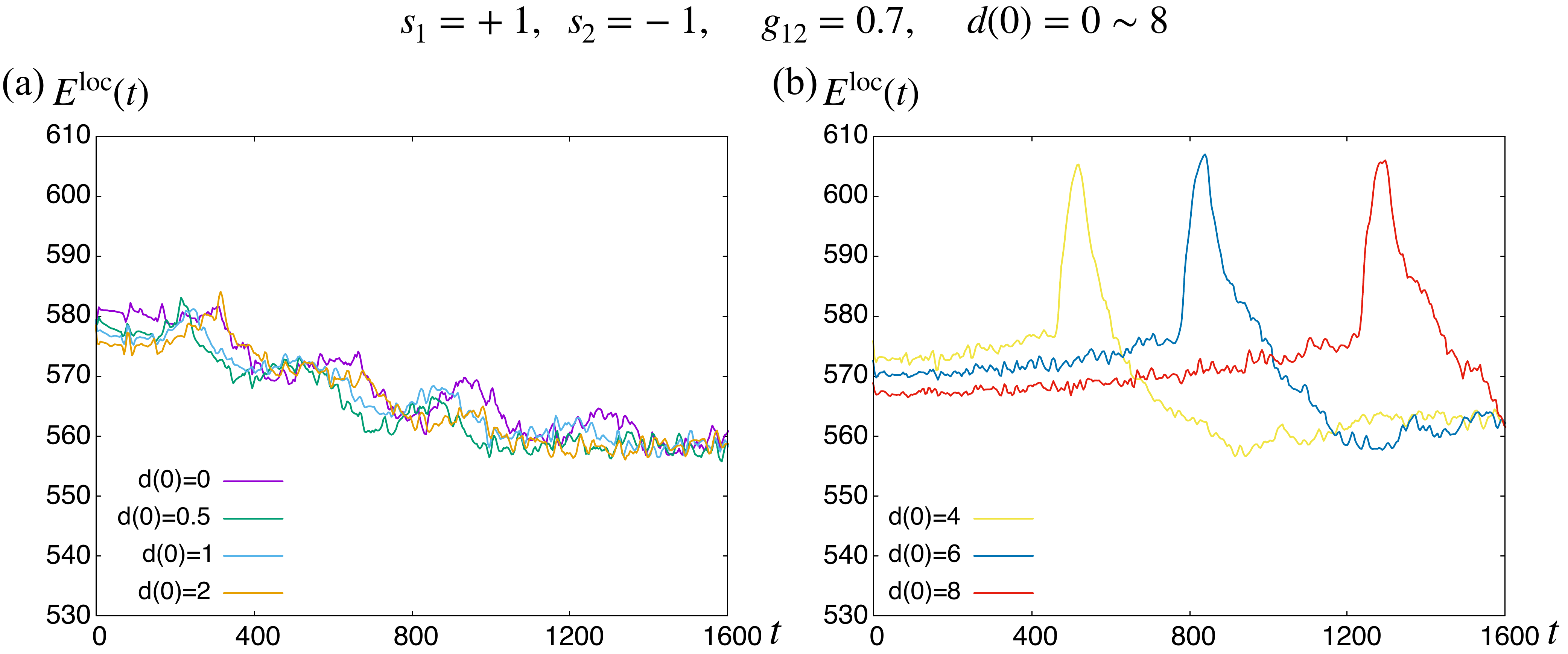}
\caption{\label{f10} 
The time development of (a) the local energy $E^{\rm loc}(t)$ for $d(0) = 0, \ 0.5, \ 1, \ 2$, and (b) $E^{\rm loc}(t)$ for $d(0) = 4, \ 6, \ 8$, for $s_{1} = + 1$, $s_{2} = -1$, and $g_{12}=0.7$.
The initial positions of the vortices were $(x_{1}(0), y_{1}(0)) = (0, d(0)/2)$ and $(x_{2}(0), y_{2}(0)) = (0, -d(0)/2)$.}
\end{figure}
\end{center}

To further realize the dynamics, we calculated the local energy $E^{\rm loc} = \int_S \epsilon^{\rm tot}({\boldsymbol r}, t) d{\boldsymbol r}$, which is defined as the total energy in the region $S$ within the circle whose center is the middle point $r^{\rm mid}$ of the two vortices with radius $r^{\rm loc} = 10$ (Fig. \ref{f9}).
This region $S$ including the vortices moves with them.
The time development of $E^{\rm loc}(t)$ is shown in Fig. \ref{f10}.
For $d(0) \leq 2$, $E^{\rm loc}(t)$ gradually decreased, reflecting the increase in $d(t)$, whereas the overlapping does not cause serious effects.  
For $d(0) > 2$, $E^{\rm loc}(t)$ shows significant peaks at $t = T_{\rm o}$, when the energy was injected into the region $S$ from the outside.

For $d(0) \leq 2$, the two vortices are co-located or close to each other.
Then, $d(t)$ hardly increases until $t = T_{\rm o}$, which is followed by the splitting.
This process is independent of $d(0)$ as long as the two vortices are nearly co-located with $d(0) \leq 2$.
For $d(0) > 2$, two vortices were separated sufficiently initially, and the overlapping occurred almost at the same distance $r_{1}(T_{\rm o}) \simeq 40 $ (Fig. \ref{f8} (d)).
Because $r_{1}(T_{\rm o}) \simeq 40 $ is close to the radius $R_{0} = 56$ of the potential, the overlapping events occurred near its boundary. 
The initial velocity of the two vortices decreased with $d(0)$, which required more time for them to reach $r_{1}(T_{\rm o}) \simeq 40 $.

%%%%%%%%%%%%%%%%%%%%
%%%%%%%%%%%%%%%%%%%%
%%%%%%%%%%%%%%%%%%%%
%%%%%%%%%%%%%%%%%%%%

\begin{center}
\begin{figure}[h]
	\includegraphics[width=15pc, height=11.25pc]{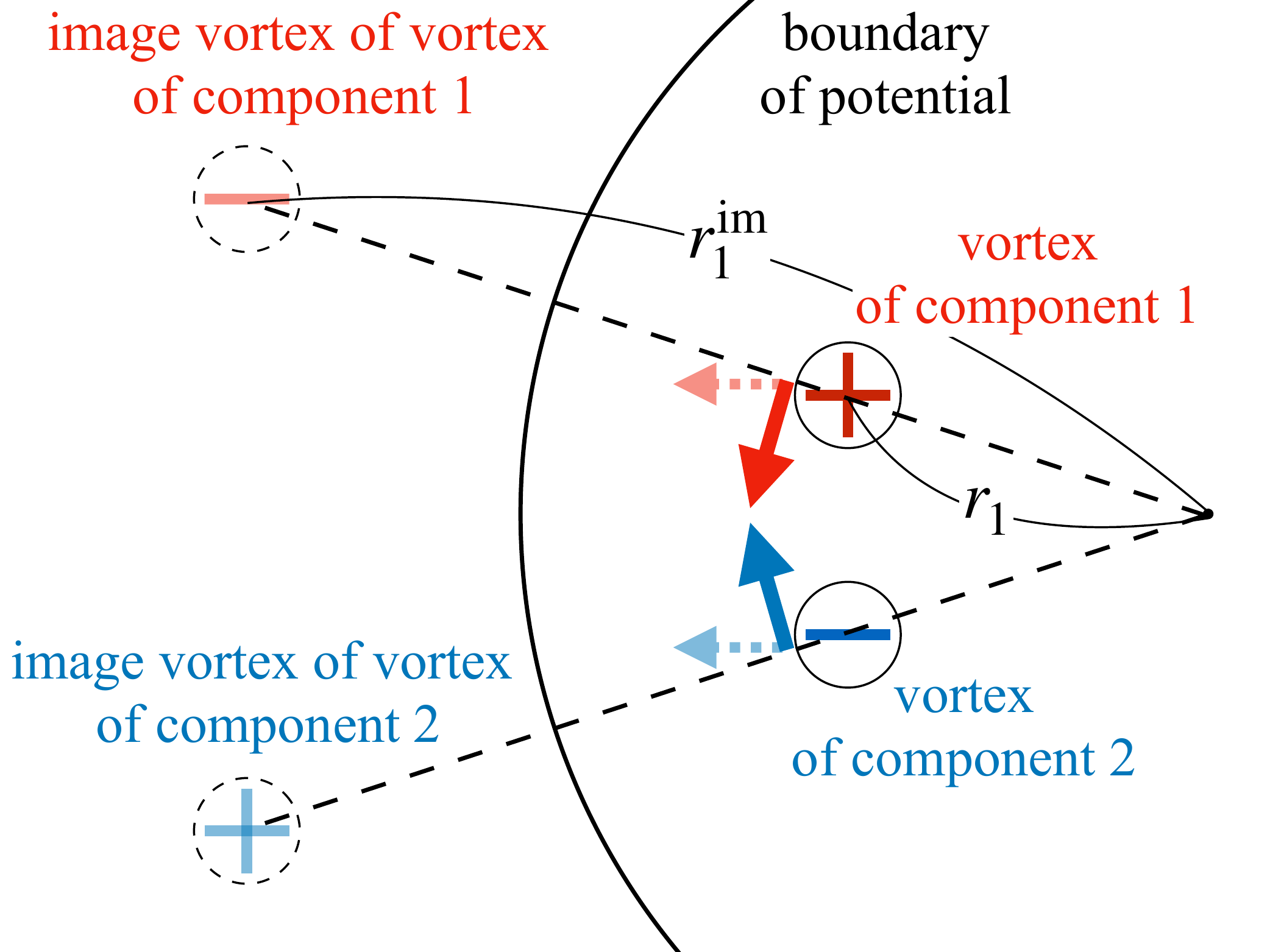}
\caption{\label{f11} 
The schema of the position of image vortex.
The solid black curved line indicates the boundary of the potential, and the pale vortices on the left side of the boundary are the image vortices.
Here, the solid arrow indicates the direction of the motion of the vortex that comes from the interaction with its image vortex, and the pale dotted arrow shows the direction calculated by Eqs.~(\ref{eq:motion}):
From these, the two vortices become closer to each other, and finally overlap.
}
\end{figure}
\end{center}

The main cause of overlapping can be attributed to the interaction between the vortex and its image vortex.
If this interaction is absent, it is expected that the two vortices keep the motion perpendicular to their dipole, and $d(t)$ does not change by Eqs.~(\ref{eq:motion}):
However, the vortex moved along the boundary owing to the effect of the interaction with their image vortex (Fig. \ref{f11}).
Here, the position vector ${\boldsymbol r}^{\rm im}_{j}$ of the image vortex can be calculated using the position vector ${\boldsymbol r}_{j}$ of the vortex as
%%%%%%%%%%
\begin{equation}
	{\boldsymbol r}^{\rm im}_{j} = \frac{R_{0}^{2}}{r_{j}^{2}}{\boldsymbol r}_{j}, \label{eq:r^im_1}
\end{equation} 
%%%%%%%%%% 
and the sign of the circulation is opposite to that of the vortex, namely, $s^{\rm im}_{j} = -s_{j}$.
Thus, these two vortices get closer to each other, and finally overlap.
For $d(0) \leq 2$, the overlapping occurs closer to the center of the potential than that for $d(0) > 2$ (Fig. \ref{f8}) since the vortices sufficiently overlap even in the initial state and the subtle interaction with the image vortex is enough to cause this overlapping.
Here, the interaction with the image vortex becomes strong with close to the boundary of the potential.

%%%%%%%%%%%%%%%%%%%%
%%%%%%%%%%%%%%%%%%%%

\begin{center}
\begin{figure}[h]
	\includegraphics[width=20pc, height=13.684pc]{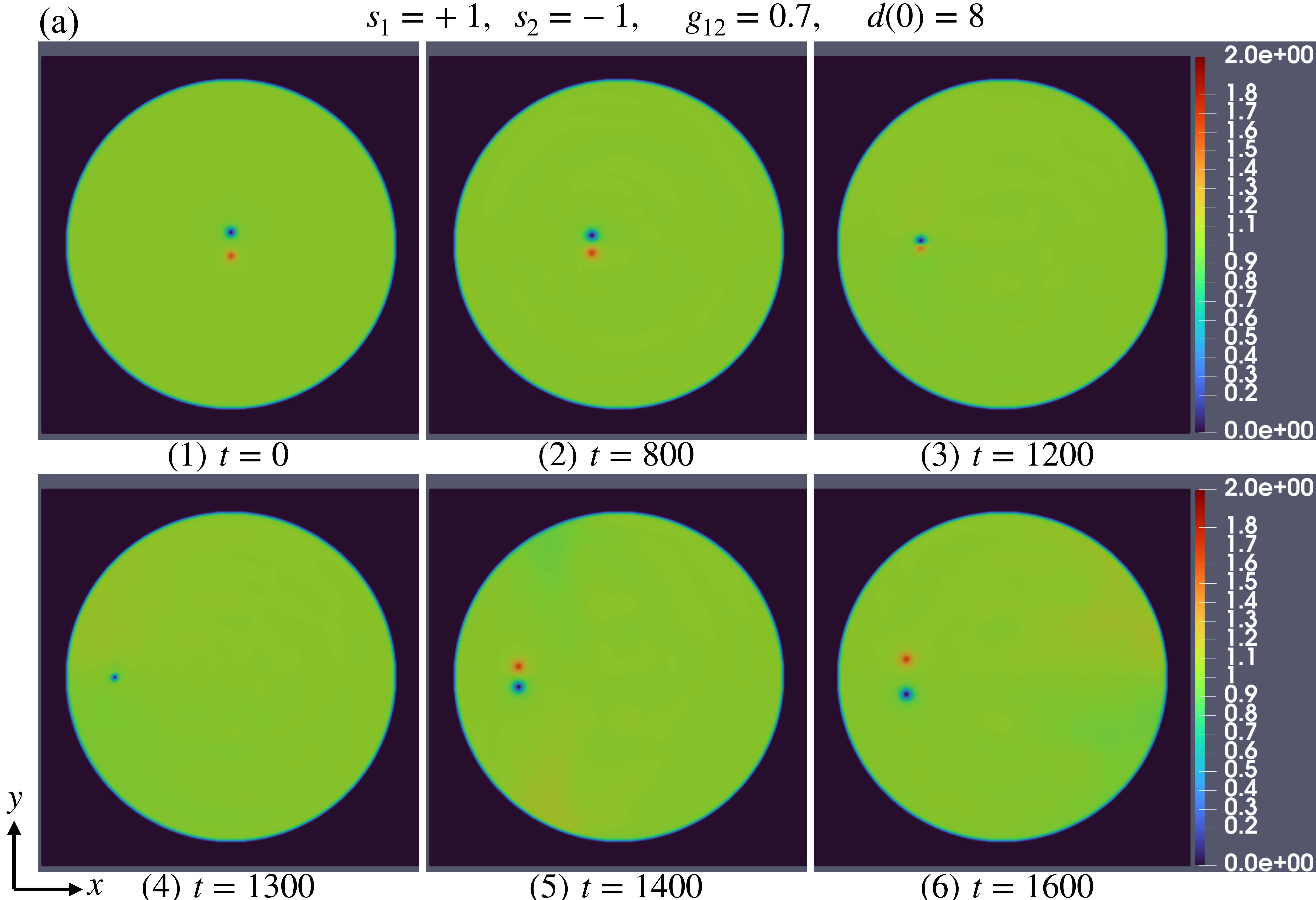}
	\includegraphics[width=20pc, height=7.6pc]{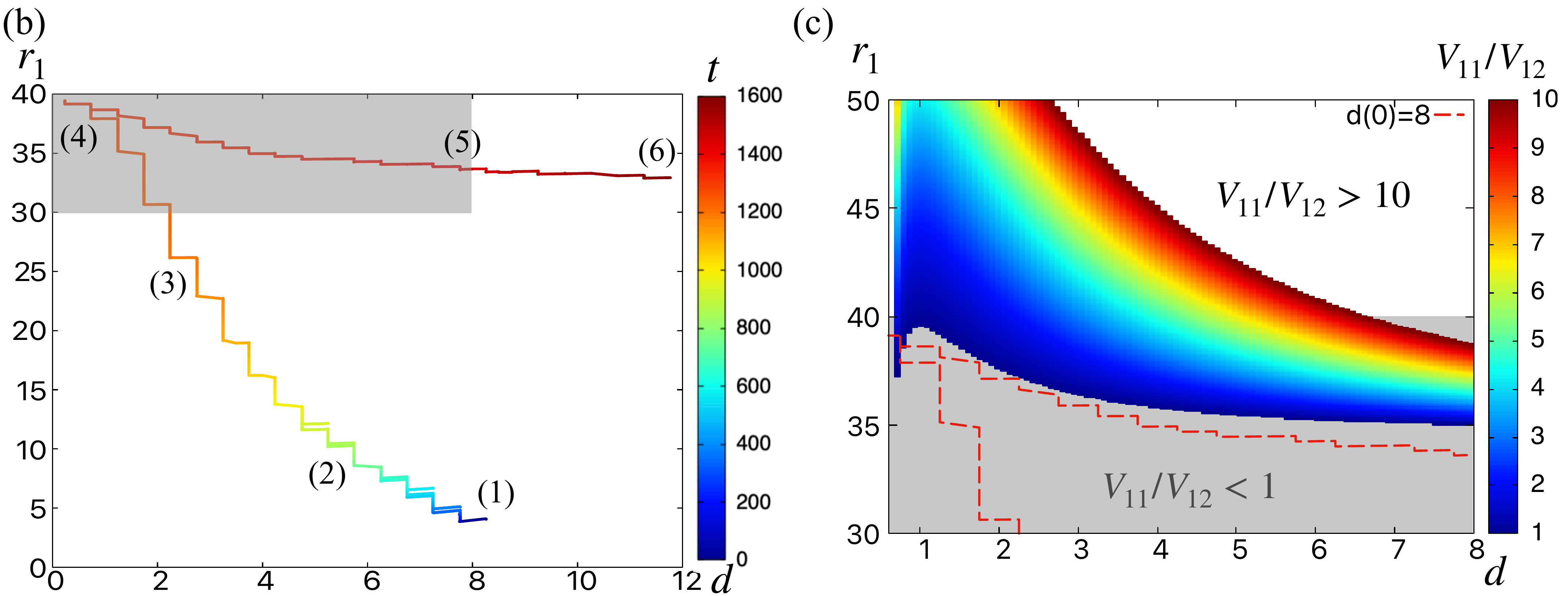}
\caption{\label{f12} 
(a) The density of component 1 for $s_{1} = + 1$, $s_{2} = -1$, $g_{12}=0.7$, and $d(0) = 8$.
(b) The time development of the $d(t)$ and $r_{1}(t)$.
Here, the color on the curve indicate the time represented by the color legend, and (1)-(6) correspond to the snapshots of (a).
(c) The ratio of the potential of the intracomponent and intercomponent interaction between vortices $V_{11}/V_{12}$ as a function of $d$ and $r_{1}$, where $r_{1\rm im}$ in the equation of the top right is calculated using Eq.~(\ref{eq:r_1_im}).
The color bar range is $1 \leq V_{11}/V_{12} \leq 10$, and the lower gray region satisfies $V_{11}/V_{12} < 1$ and the upper white region satisfies $V_{11}/V_{12} > 10$.
The gray region of (b) corresponds to the gray region of (c), and the red dashed line shows the same line as the colored line in (b).}
\end{figure}
\end{center}

We now reveal the details of the mechanism of the overlapping assisted by the image vortices.
Figure \ref{f12} (a) illustrates the density of component 1 for $s_{1} = + 1$, $s_{2} = -1$, $g_{12}=0.7$, and $d(0) = 8$.
Figure. \ref{f12} (b) shows the time development of $d$ and $r_{1}$. 
The two vortices move to the same direction (Fig. \ref{f12}, (1) - (3)), then overlapped [(4)], and finally reversed [(5) and (6)].

We attempt to estimate the region where the effect of the image vortex becomes significant for $d(0) > 2$.
From Fig. \ref{f11} and Eq. (\ref{eq:r^im_1}), the distance between the vortex and its image vortex is:
%%%%%%%%%%
\begin{equation}
	r_{j {\rm im}} = r^{\rm im}_{j} - r_{j} = \frac{R_{0}^{2}}{r_{j}} - r_{j}. \label{eq:r_1_im}
\end{equation} 
%%%%%%%%%% 
Using this, we calculate the ratio of the intracomponent interaction $V_{11}$ between the vortices and the intercomponent interaction $V_{12}$:
%%%%%%%%%%
\begin{equation}
	\frac{V_{11}}{V_{12}} = 2\frac{1 - (g_{12})^{2}}{g_{12}}d^{2} \frac{\ln(R/r_{1{\rm im}})}{\ln(d/{2\xi_{\rm cut}})}. \label{eq:eintra_einter}
\end{equation} 
%%%%%%%%%%
In Fig. \ref{f12} (c), the dependence of $V_{11}/V_{12}$ on the $d$ and $r_{1}$ is shown; the colored region satisfies $1 \leq V_{11}/V_{12} \leq 10$, where the intracomponent interaction is larger than the intercomponent one.
The lower gray region satisfies $V_{11}/V_{12} < 1$. 
We observed that the overlapping (Fig. \ref{f12} (4)) and the subsequent vortex motion [(5) and (6)] took place along with the contours of the colored region in (c), especially the curve of $V_{11}/V_{12} \sim 1$.

%%%%%%%%%%%%%%%%%%%%
%%%%%%%%%%%%%%%%%%%%

Figure \ref{f12} (c) implies the difference of the mechanism of overlapping between $d(0) > 2$ and $d(0) \leq 2$.
In the former, overlapping occurred at $r_{1}(T_{\rm o}) \simeq 40$ inside the colored region spread for $r_{1} \gtrsim 35$ (Fig. \ref{f8} (d)), thus, the intracomponent interaction with the image vortex causes overlapping for $d(0) > 2$.
In the latter, however, overlapping occurred around $r_{1}(T_{\rm o}) \simeq 30$ outside the colored region (Fig. \ref{f8} (c)), being not driven by the image vortex mainly.

%%%%%%%%%%%%%%%%%%%%%%%%%%%%%%%%%%%%%%%%%%%%%%%%%%%%%%%%%%%%%%%%%%%%%%%%%%%%%%%%%%%%

\section{Conclusions} \label{sec:Conclusions}

We examined the characteristic dynamics of two vortices belonging to different components confined by a circular box potential when they are initially co-located or close to each other. 
We then revealed three characteristic results.

First is the splitting of two vortices from the initial co-located state.
This is owing to the dynamical instability of the co-located state, which is revealed by the BdG analysis.
In Fig. \ref{f1}, the diagram of the instability is shown for both $s_{1} = s_{2} = +1$ state and $s_{1} = +1$, $s_{2} = -1$ state.
We also confirmed the instability of the two vortices in the both states using the GP equation, which is consistent with the diagram.

The second is the dynamics after the splitting with $s_{1} = s_{2} = +1$.
In this case, the two vortices generated a circular motion around their center.
If the two vortices are co-located initially, they split, then the distance between them and the angular velocity changes (Fig. \ref{f5} (a)).
The angular velocity can be obtained as a function of the distance between the two vortices (Eq. (\ref{eq:phi_dif})), derived from the equations of motion (Eqs.~(\ref{eq:motion})).
However, if they are separated sufficiently initially, the changes in the distance and the angular velocity became very small (Fig. \ref{f5} (c)).

The third is the dynamics after the splitting with $s_{1} = +1$ and $s_{2} = -1$.
In this case, the two vortices initially moved in the same direction, which is perpendicular to their dipole.
They then overlapped for a moment during the dynamics.
We can categorize this overlapping into two types, $d(0) \leq 2$ or $d(0) > 2$ by the position where the overlapping occurs (Fig. \ref{f8}) and the time development of the local energy (Fig. \ref{f10}).
Here, the overlap for $d(0) \leq 2$ occurs by a slight contribution from the interaction with an image vortex. 
Then, the local energy hardly increases through this process.
However, the overlap for $d(0) > 2$ comes from the sufficient interaction with its image vortex (Fig. \ref{f11}), and thus, the local energy shows significant peaks and this overlapping occurs closer to the boundary of the potential than that of former (Fig. \ref{f12}).

Studying these elementary dynamics is important for realizing the multicomponent QT because the moving obstacle potential emits the co-located vortices \cite{Mithun21}, and the co-located vortices can also split through the dynamics (Appendix).
This multicomponent QT has been experimentally studied by a system confined by the trapping potential, and thus, the effect of the boundary on the dynamics cannot be neglected.
Because there are many vortices in multicomponent QT, the process of the splitting and the dynamics of vortices is also affected by other vortices. 

Finally, we comment about the experimental realization and the remaining problems.
The box potential has already been used in experiments \cite{Gaunt13, Navon16}, and the initial condition can be created by the phase imprinting \cite{Matthews99, Leanhardt02}; thus, this study can be reproduced in an experiment.
We extensively focused on the dynamics after the splitting, but there are several unresolved problems.
One of them is to examine what determines the direction of the dipole of two vortices, from which the direction of the motion after the splitting is determined.
Another is to explain what makes the change of the distance between vortices.
In a previous study \cite{Kasamatsu16}, the equations of motion and the asymptotic form of the interaction between two vortices (Eqs. (\ref{eq:motion}) and (\ref{eq:V_12})) were proposed, in which the distance between the two vortices is conserved during the dynamics.
in contrast, the distance changes in the simulations of the GP equations in both our study and the previous one \cite{Kasamatsu16}.
The change in the distance after splitting could be caused by the interaction between the vortices and phonons, which originate from the excess energy and are emitted when the two vortices split.
The detailed mechanism of this is of merit for further study.

%%%%%%%%%%%%%%%%%%%%%%%%%%%%%%%%%%%%%%%%%%%%%%%%%%%%%%%%%%%%%%%%%%%%%%%%%%%%%%%%%%%%

\section{acknowledgment}\label{sec:acknowledgment}

This work was supported by a Grant-in-Aid for JSPS Fellows (grant nos. 20J14459, 18K03472, and JP20H01855).

%%%%%%%%%%%%%%%%%%%%%%%%%%%%%%%%%%%%%%%%%%%%%%%%%%%%%%%%%%%%%%%%%%%%%%%%%%%%%%%%%%%%

\section{Appendix}\label{sec:appendix}

We introduce our preliminary calculations on the two-component turbulence caused by the obstacle potential in a uniform flow.
The motivation of this study was attributed to this calculation.

We assumed a uniform system of miscible two-component BECs with a stationary Gaussian obstacle potential and uniform flow.
Then, the GP equation of this system is
%%%%%%%%%%
\begin{eqnarray}
	\imath\hbar\frac{\partial}{\partial t}\psi_{j} &=& \left \{-\frac{\hbar^{2}}{2m_{j}}\nabla^{2} + V^{\rm obs} + \displaystyle\sum_{j'=1,2} g_{jj'}|\psi_{j'}|^{2} - \mu_{j} \right \} \psi_{j} \nonumber \\
	&-& ({\boldsymbol v}_{\rm flow} \cdot {\boldsymbol p}) \psi_{j}. \label{eq:GPEs_flow}
\end{eqnarray} 
%%%%%%%%%%
The velocity of the flow ${\boldsymbol v}_{\rm flow}$ is normalized by $\xi / \tau = \sqrt{gn/m}$.
The Gaussian obstacle potential is 
%%%%%%%%%%
\begin{equation}
	V^{\rm obs}(x, y) = V^{\rm obs}_{0} \exp\left(-\frac{(x - x^{\rm obs})^{2} + (y - y^{\rm obs})^{2}}{(d^{\rm obs})^{2}} \right), \label{eq:obs_pot}
\end{equation} 
%%%%%%%%%%
and the boundary condition of the system is periodic.

%%%%%%%%%%%%%%%%%%%%
\begin{center}
\begin{figure}[h]
	\includegraphics[width=20pc, height=11.5pc]{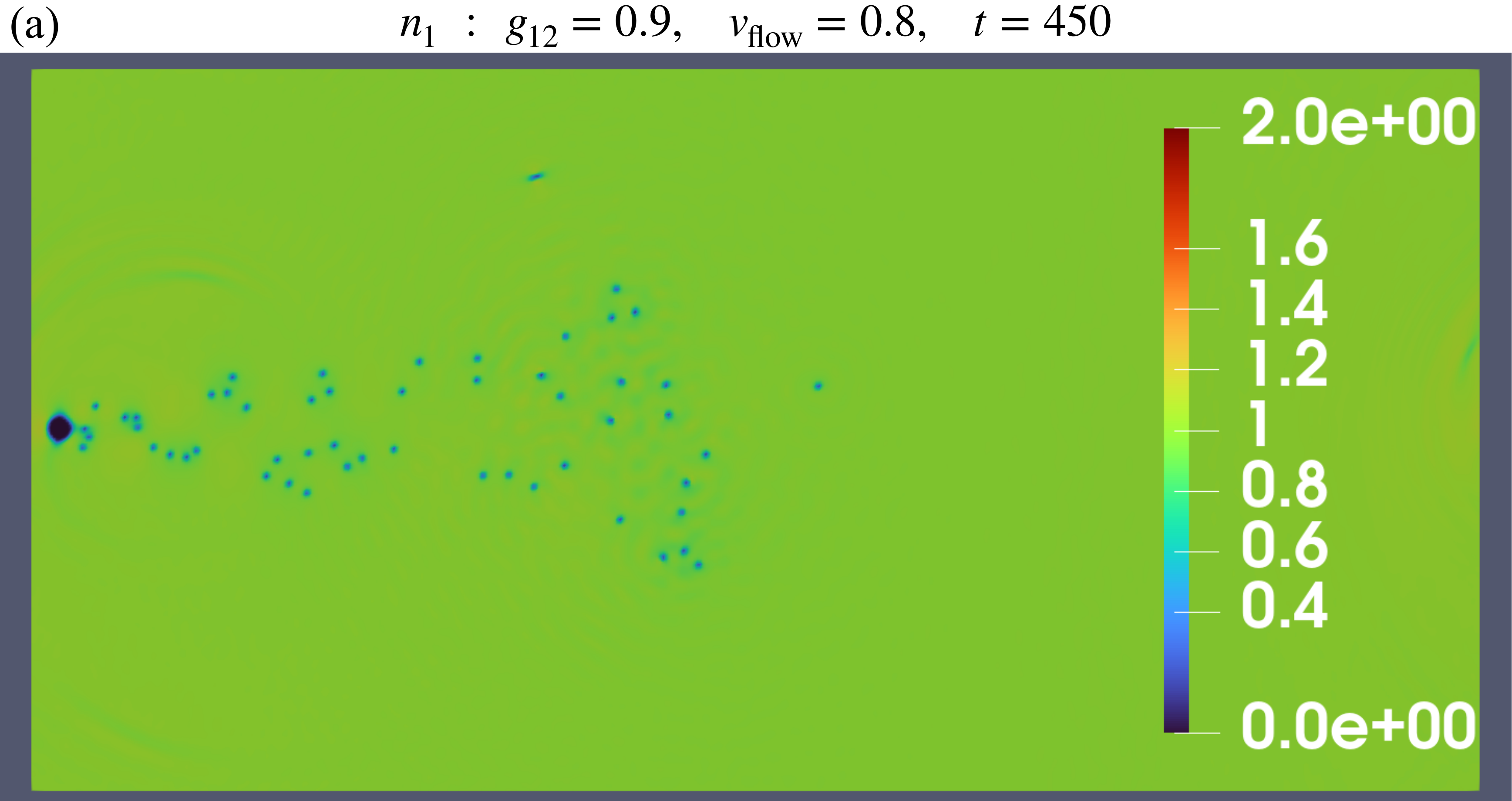}
	\includegraphics[width=20pc, height=11.5pc]{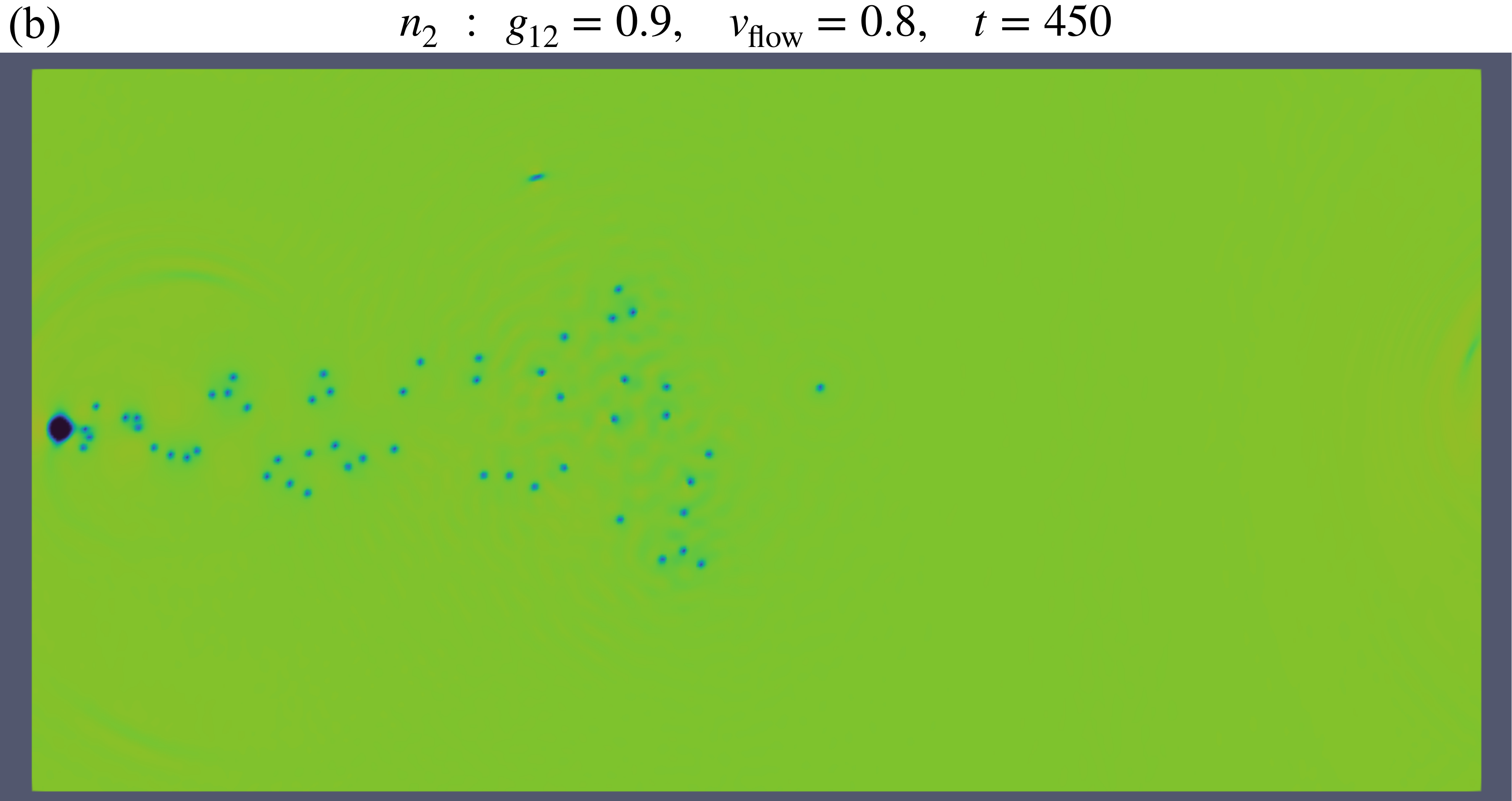}
	\includegraphics[width=20pc, height=11.5pc]{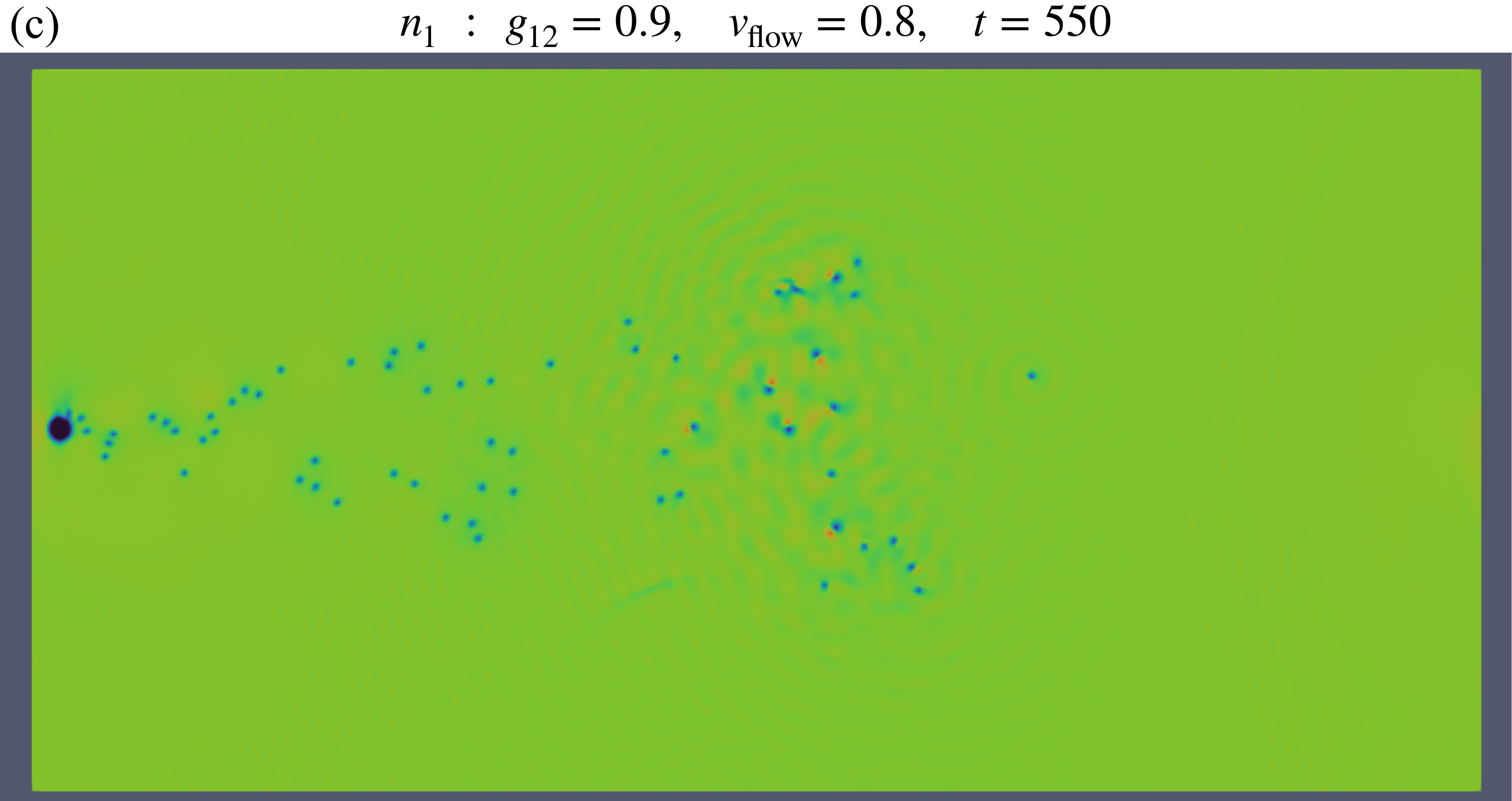}
	\includegraphics[width=20pc, height=11.5pc]{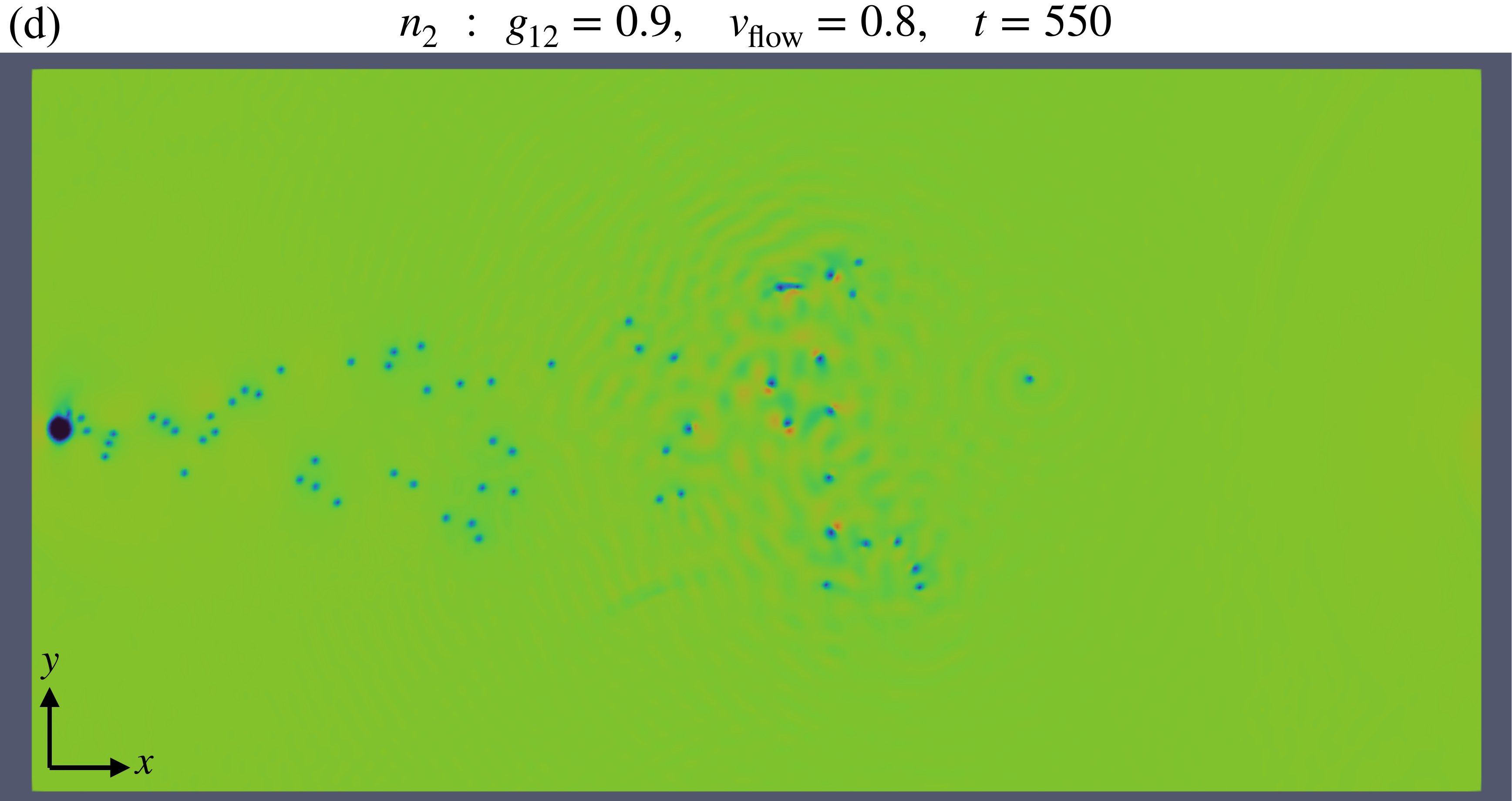}
\caption{\label{f1a} 
The densities of (a) and (c) component 1, and (b) and (d) component 2 for $g_{12} = 0.9$.
Here, (a) and (b) show that at $t=450$, and (c) and (d) show that at $t=550$.
In these calculations, the calculation range is $-256\xi \leq x \leq 256\xi$, $-64\xi \leq y \leq 64\xi$, and each parameter is $V^{\rm obs} = 100 gn$, $x^{\rm obs} = -240\xi$, $y^{\rm obs} = 0\xi$, and $d^{\rm obs} = 2\sqrt{2}\xi$.
The left large low-density circle is the Gaussian obstacle potential, and the uniform flow is induced from left to right, whose velocity is $v_{\rm flow} = 0.8$, and this velocity is normalized by $\xi / \tau$.}
\end{figure}
\end{center}
%%%%%%%%%%%%%%%%%%%%

Figure \ref{f1a} shows the density distribution for $g_{12} = 0.9$ and $v_{\rm flow} = 0.8$ which flows from left to right.
Then, the vortices of both components emitted by the potential are co-located for a long time $t \lesssim 450$ (Fig. \ref{f1a} (a) and (b)).
This is consistent with the results of the binary QT in a harmonic potential \cite{Mithun21}.
Some co-located vortices start to split downstream (Fig. \ref{f1a} (c) and (d), $t = 550$), and the time for splitting depends on $g_{12}$.

Until the splitting, the dynamics of the QT is similar to that of the one-component QT, whereas the splitting makes the dynamics more different.
Thus, we should know the splitting and the dynamics after that to realize binary QT.

%%%%%%%%%%%%%%%%%%%%%%%%%%%%%%%%%%%%%%%%%%%%%%%%%%%%%%%%%%%%%%%%%%%%%%%%%%%%%%%%%%%%

\bibliographystyle{jpsj}
\bibliography{book,paper}

%\begin{thebibliography}{9}
%\bibitem{jpsj} The abbreviation for JPSJ must be ``J. Phys. Soc. Jpn." \note{in the reference list}.
%\bibitem{instructions} More abbreviations of journal titles are listed in ``Instructions for Preparation of Manuscript".
%\bibitem{Errata} Errata should be listed under the same reference number. 
%\end{thebibliography}

\end{document}